\documentclass[aps,prd,showpacs,floatfix,twocolumn,
nofootinbib,preprintnumbers,
10pt]{revtex4-2}
\makeatletter
\def\p@subsection{}
\makeatother

\usepackage{graphicx,amsmath,amssymb,lipsum,bm}

\usepackage[dvipsnames]{xcolor}
\definecolor{xlinkcolor}{rgb}{0.7752941176470588, 0.22078431372549023, 0.2262745098039215}
\definecolor{BrickRed}{rgb}{0.7752941176470588, 0.22078431372549023, 0.2262745098039215}
\definecolor{xlinkcolor}{HTML}{1c1e94}
\usepackage{booktabs} 
\usepackage[normalem]{ulem}
\usepackage[utf8]{inputenc} 
\usepackage{mathtools}
\usepackage{aux_files/aas_macros}
\usepackage{xspace}

\usepackage{ulem}
\usepackage{diagbox}

\usepackage[dvipsnames]{xcolor}
\usepackage{mathrsfs}

\usepackage[colorlinks=true,citecolor=xlinkcolor,linkcolor=xlinkcolor,urlcolor=xlinkcolor, backref=false,pdfborder={0 0 0}]{hyperref}

\newcommand{\beqa}{\begin{eqnarray}}
\newcommand{\eeqa}{\end{eqnarray}}

\newcommand{\be}{\begin{equation}}
\newcommand{\ee}{\end{equation}}
\newcommand{\beq}{\begin{equation}}
\newcommand{\eeq}{\end{equation}}

\newcommand{\bseq}{\begin{subequations}}
\newcommand{\eseq}{\end{subequations}}

\def\ltsima{$\; \buildrel < \over \sim \;$\xspace}
\def\gtsima{$\; \buildrel > \over \sim \;$\xspace}
\def\simlt{\lower.5ex\hbox{\ltsima}}
\def\simgt{\lower.5ex\hbox{\gtsima}}

\usepackage{multirow}

\def\gsim{\raise0.3ex\hbox{$\;>$\kern-0.75em\raise-1.1ex\hbox{$\sim\;$}}}
\def\lsim{\raise0.3ex\hbox{$\;<$\kern-0.75em\raise-1.1ex\hbox{$\sim\;$}}}

\def\beqn#1{\begin{equation}\label{#1}}
\def\eeqn{\end{equation}}

\def\beqa#1{\begin{eqnarray}\label{#1}}
\def\eeqa{\end{eqnarray}}

\def\Z2{$\mathcal{Z_2}$}

\newcommand {\ignore}[1]{}

\renewcommand{\arraystretch}{1.3} 

\DeclareRobustCommand{\ion}[2]{%
\relax\ifmmode
\ifx\testbx\f@series
{\bm{#1\,\mathsc{#2}}}\else
{\mathrm{#1\,\mathsc{#2}}}\fi
\else\textup{#1\,{\mdseries\textsc{#2}}}%
\fi}

\newcommand{\MIT}{Center for Theoretical Physics -- a Leinweber Institute, Massachusetts Institute of Technology, Cambridge, MA 02139, USA}
\newcommand{\IAIFI}{The NSF AI Institute for Artificial Intelligence and Fundamental Interactions, Cambridge, MA 02139, USA}

\begin{document}

\preprint{MIT-CTP/6048}

\title{
Cosmological Concordance in 
an Especially Opaque
Universe: \\ A Tentative Cosmological 
Detection of Physical  
Neutrino Mass in $\Lambda$CDM 
}
\author{James M. Sullivan}
\email{jms3@mit.edu}\thanks{Brinson Prize Fellow}
\author{Roger de Belsunce}
\author{Mikhail M. Ivanov}\email{ivanov99@mit.edu}
\affiliation{\MIT}\affiliation{\IAIFI}

\date{\today}

\begin{abstract}
The measurement of the 
sum of neutrino masses is among the primary promises of 
precision cosmology, achievable by combining complementary early- and late-Universe probes. 
However, these datasets currently exhibit mild-to-strong disagreements within $\Lambda$CDM and its simplest extensions, giving rise to multiple tensions, including the Hubble tension, the preference for ``negative'' neutrino mass, 
and indications of evolving dark energy.
It has recently been shown that 
these tensions can be 
alleviated by adopting a higher value of the optical depth to reionization parameter, $\tau$, when large-scale CMB polarization data are ignored.
We extend this proposal and show that a single, especially high prior on 
$\tau = 0.11 \pm 0.006$ simultaneously addresses all three of these tensions, 
thereby significantly  reducing the need for new physics beyond $\Lambda$CDM. We determine the ``concordance'' value
of 
$\tau$
by requiring (i)
the physical masses of neutrinos
to be positive, and 
(ii) consistency of the $\Lambda$CDM-based Hubble constant, $H_0$, inferred from the cosmic microwave background (CMB) and that preferred by the Dark Energy Spectroscopic Instrument (DESI) baryon acoustic oscillation (BAO) and full-shape galaxy-clustering measurements. Within this high-$\tau$ Universe, we obtain the first $2\sigma$ detection of a positive neutrino mass, $\Sigma m_{\nu} = 0.10^{+0.04}_{-0.05}$~eV at 68\% C.L., while restoring cosmological concordance between datasets within $\Lambda$CDM. 
In particular, the resulting cosmology predicts low-redshift distance measurements consistent with DESI BAO observations and raises the inferred value of $H_0$ into agreement with the CMB-independent determination from DESI. Furthermore, the inferred dark-energy equation-of-state parameters become consistent with a cosmological constant, both with and without supernova data.
We find that increasing $\tau$ reduces the Hubble tension with the SH0ES measurement to $4.4\sigma$, while 
remaining fully consistent with the
Chicago-Carnegie Hubble Program (CCHP) measurement.
The concordance power of our $\tau$ prior further motivates new measurements of 
$\tau$, for example through large angular-scale CMB polarization observations with the \textit{LiteBIRD}, CLASS, or proposed PICO experiments.
\end{abstract}

\maketitle

\section{Introduction and Summary \label{sec:intro}}

The last decade in cosmology has seen the emergence of multiple tensions between precision datasets.
Key cosmological data in the early and late Universe have produced mild, moderate, and strong disagreements between the parameters of the standard $\Lambda$CDM model and its simplest extensions.
Three of these tensions stand out,
with the most prominent being the \emph{Hubble tension}, the long-standing disagreement between direct and indirect determinations of the Hubble constant; see~\cite{DiValentino2021:H0review,Abdalla:2022yfr} for reviews.
Nominally, the cosmic microwave background (CMB)-inferred value, $H_0=67.4\pm 0.5$ km/s/Mpc \cite{planck18}, is in a $\approx 6\sigma$ tension with the most recent Cepheid-calibrated Type Ia supernovae measurement, $H_0=73.17\pm 0.86$ km/s/Mpc~\cite{Breuval:2024lsv} (superseding the previous result $H_0=73.04\pm 1.04$ km/s/Mpc \cite{Riess:2021jrx}).
Despite being a problem for over a decade, this local-Universe $H_0$ tension still does not have a satisfactory solution, neither in terms of compelling new physics models, nor in terms of known sources of measurement systematics~\cite[see, e.g.,~][]{Efstathiou:2021ocp}.
The value of the Hubble constant inferred by the combination of DESI full-shape measurements and CMB lensing is also somewhat larger than the CMB-inferred value \cite{Ivanov2026:desi_dr1_4_kitchen_sink,Maus2025:desi_lensing_xcorr}.
The second tension is the cosmological preference for ``negative'' neutrino mass, driven by the CMB lensing anomaly \cite{planck18, planck_lens_18}. The third is the preference for evolving dark energy suggested by the Dark Energy Spectroscopic Instrument (DESI;~\cite{desibao2024, AbdulKarim2025:desi_dr2_bao}).

In this work we show that a single, especially high prior on the optical depth to reionization, $\tau = 0.11 \pm 0.006$, simultaneously alleviates all three of these tensions within $\Lambda$CDM when neglecting large-angle polarization data from \textit{Planck}, thereby largely removing the need for new physics. 
Most strikingly, it converts the preference for ``negative'' neutrino mass into a tentative $2\sigma$ detection of a positive, physical neutrino mass.
This prior does not agree with low-$\ell$ \textit{Planck} $EE$ polarization data, but
it crucially allows us 
to determine whether it is possible to find evidence for physical neutrino mass with the precision of current data within $\Lambda$CDM if the central value of $\tau$ departs significantly from the value inferred from this CMB data. 
We find that, indeed, this is the case.
This prior also removes evidence for dynamical dark energy, and substantially alleviates the Hubble tension with respect to $H_0$ preferred by DESI data, while somewhat reducing its significance with respect to the local SH0ES measurement. We describe all the three tensions in the following. 

The most central tension to this work is the cosmological preference for ``negative'' neutrino mass.
Its origin lies in the CMB anisotropy spectrum measured by the \textit{Planck} satellite, which shows a mild disagreement between the small-scale and large-scale parts of the spectrum, producing a preference for excess gravitational lensing ($A_\mathrm{L}>1$) relative to the $\Lambda$CDM prediction.
This can be recast as a tension between the primary-CMB-predicted matter fluctuation amplitude, $\sigma_8$, and its direct measurement from both the smoothing of the acoustic peaks and CMB lensing reconstruction~\cite{planck18, planck_lens_18}.
This excess lensing translates into especially strong constraints on the neutrino mass sum, with a significant preference for zero (or ``negative'') neutrino masses~\cite{Craig2024:neg_nu,LoverdeWeiner2024:BAO,Green:2024xbb,Elbers2025:neg_neutrinos,Graham2026:neg_nu_G2,Herold:2024nvk};
a preference which is especially dramatic due to degeneracy breaking when lensing reconstruction information is combined with geometric constraints from the CMB angular scale of the sound horizon \cite{LoverdeWeiner2024:BAO}
At face value, the preference for ``negative'' neutrino masses is in conflict with particle-physics oscillation experiments, suggesting new physics, e.g. neutrino decays~\cite{Craig2024:neg_nu,Chacko:2020hmh}.

The third, especially timely, tension is the preference for time evolution in the equation of state of dark energy, suggested by measurements of the baryonic acoustic oscillation (BAO) feature observed by DESI~\cite{DESI:2016, desibao2024, AbdulKarim2025:desi_dr2_bao} in combination with CMB data~\cite{desibao2024,AbdulKarim2025:desi_dr2_bao}.
In $\Lambda$CDM, DESI BAO favors lower distances to observed galaxies than those predicted by the \textit{Planck} primary CMB, which leads to a lower value of the matter density, $\Omega_m$.
Their combination leads to an additional preference for zero (or ``negative'') neutrino mass, as well as non-negligible evidence for evolving dark energy (e.g., $w_0w_a$), especially when combined with uncalibrated Type Ia supernovae magnitudes.

Three aspects complicate a joint resolution of this triple trouble of tensions.
First, the dark-energy and ``negative'' neutrino-mass tensions are partly correlated, as the evidence for ``negative'' neutrino mass reduces in the $w_0w_a$CDM background~\cite{Green:2024xbb}; yet it does not go away completely, suggesting that dynamical dark energy alone cannot account for the ``negative'' neutrino-mass inference.
This is consistent with the view presented in Ref.~\cite{Weiner2026}, in which the neutrino mass tension is phrased in terms of a disagreement of high-redshift distances.
Second, while the preference for $w_0w_a$CDM cosmology is somewhat reduced by marginalizing over $A_\mathrm{L}$, non-trivial values of $(w_0,w_a)$ are still preferred over the cosmological-constant values by $\approx 2\sigma$~\cite{Peng:2025nez}.
Third, the Hubble tension resurfaces across these datasets: the inferred value of $H_0$ is low from all of them compared to certain local measurements \cite{Riess:2021jrx,Freedman2025:CCHP,Breuval:2024lsv}
, and is pushed to even lower values in an evolving dark energy model 
(by $3\sigma$,
see Tab.~V of Ref.~\cite{AbdulKarim2025:desi_dr2_bao}).

Taken at face value, the above tensions require multiple new physics sectors to restore concordance in cosmology.
The DESI evidence for evolving dark energy alone requires a complex dark sector with dark-matter--dark-energy interactions.
Accounting for ``negative'' neutrino masses requires, e.g., new interactions in the neutrino sector, 
while the Hubble tension requires new degrees of freedom during recombination, such as early dark energy, neutrino self-interactions, or primordial magnetic fields; see Ref.~\cite{Abdalla:2022yfr,h0_olympics} for a review of new physics models to address the cosmological tensions.
Rather than invoking new-physics sectors, we postulate that the optical depth to reionization $\tau$ is largely responsible for the apparent discord.

The optical depth is currently determined from large-angular-scale CMB $EE$ polarization to be $\tau \approx 0.06 \pm 0.005$ from \textit{Planck}~\cite{planck18, deBelsunce:2021mec}.
Measurements of $\tau$ have evolved significantly over the past 20 years: \emph{WMAP} reported $\tau = 0.089 \pm 0.014$~\cite{2013ApJS..208...20B}, which was refined by the \textit{Planck} 2018 legacy release using the map-making algorithm \texttt{SRoll1}~\cite{Planck:2016kqe} to $\tau_{EE}^{\texttt{SRoll1}} = 0.0506 \pm 0.0086$~\cite{planck18}.
A re-analysis based on the improved \texttt{SRoll2} map-making algorithm~\cite{Delouis:2019bub}, including temperature and polarization quadratic maximum likelihood cross-spectra, yielded $\tau^{\texttt{SRoll2}}_{\mathrm{TTTEEE}}=0.0627^{+0.0050}_{-0.0058}$~\cite{deBelsunce:2021mec, deBelsunce:2022mwk} (see, e.g.,~\cite{2013ApJS..208...20B, Lattanzi:2016dzq, Natale:2020owc, Genesini:2026lmg, CLASS:2025khf, Kageura:2026ryq} for additional measurements).
These direct determinations all lie well above the Gunn-Peterson lower bound of $\tau \gtrsim 0.04$, which gives strong astrophysical evidence that the intergalactic medium was highly ionized by $z_{\rm re} \sim 6.5$~\cite{1965ApJ...142.1633G}; they are, however, substantially lower than the value $\tau\approx 0.10-0.11$ that, as we argue below, current cosmological datasets favor (excluding low-$\ell$ EE polarization data, see Sec.~\ref{sec:data}).

A powerful, CMB-independent handle on $H_0$ comes from the full-shape (FS) analysis of the DESI  galaxy power spectrum and bispectrum, which within $\Lambda$CDM yields $H_0=69.08\pm 0.37$ km/s/Mpc~\cite{Ivanov2026:desi_dr1_4_kitchen_sink}, about $2\sigma$ higher than the \textit{Planck} 2018 best-fit value.
Crucially, this full-shape broadband measurement is only weakly affected by the geometric degeneracy~\cite{Ivanov:2019pdj,Chudaykin:2025aux,Chudaykin:2025lww,Ivanov2026:desi_dr1_4_kitchen_sink}, which makes it more robust with respect to uncertainties in other cosmological parameters -- such as the baryon and dark matter densities -- than the CMB.
We adopt this late-Universe value as the concordance target for $H_0$.

It was noticed recently in Refs.~\cite{Allali2026:reion_allali_H0_1, Allali2025:tau_H0_lowEE,LoverdeWeiner2024:BAO,Sailer2026:hightau,Jhaveri2025:hightau,act_exts_calabrese} that both the DESI distance tension and the ``negative'' neutrino masses can be alleviated by increasing the value of the optical depth to reionization, $\tau$, to $\approx 0.09$\footnote{See also Ref.~\cite{Giar2024:freetau_lowEE} for an earlier discussion of the consequences of removing low-$\ell$ EE data.}.
The $\tau=0.09 \pm 0.01$ Universe considered by Refs.~\cite{Sailer2026:hightau,Jhaveri2025:hightau,act_exts_calabrese}, however, still exhibited a $\approx 2\sigma$ preference for non-trivial values of $(w_0,w_a)$ and did not yield a preference for positive neutrino masses.
In addition, CMB data combined with a fixed value of $\tau=0.09$ predict a somewhat low value of $H_0=67.94\pm 0.44$ km/s/Mpc, which is still in tension with SH0ES and in mild $\simeq 2\sigma$ tension with the DESI full-shape anchor introduced above~\cite{Ivanov2026:desi_dr1_4_kitchen_sink}.
This indicates that the value of $\tau$ required to restore full cosmological concordance is larger than $0.09$.

We can estimate this ``concordance'' value of $\tau$ directly by requiring consistency between the $H_0$ inferences from CMB and large-scale-structure datasets within $\Lambda$CDM.
Using the correlation between $\tau$ and $H_0$ from the \textit{Planck} 2018 base-$\Lambda$CDM chains,
\be
\label{eq:deltatauH}
\Delta \tau
\approx \frac{\Delta H_0}{33\,\text{km/s/Mpc}}\,,
\ee
reaching the DESI full-shape value $H_0=69.1$ km/s/Mpc requires $\tau\approx 0.11$. 
This value is in strong tension with the Planck-low-$\ell$ EE constraint, but is interesting to explore as a possible mechanism to raise the value of $H_0$.
Motivated by this, in what follows we replace the standard low-$\ell$ EE likelihood with a Gaussian prior centered at $\tau_*=0.11$, with a width equal to that of the standard low-$\ell$ EE likelihood, $0.006$. We refer to this as the high-$\tau$ prior, or a high-$\tau$ Universe. 
Reassuringly, a direct determination of $\tau$ from current cosmological datasets (see Refs.~\cite{Jhaveri2025:hightau,Sailer2026:hightau}, Sec.~\ref{subsec:fullshape}) independently lands near $\tau\approx 0.10-0.11$, corroborating this choice.

\begin{figure}
    \centering
    \includegraphics[width=\linewidth]{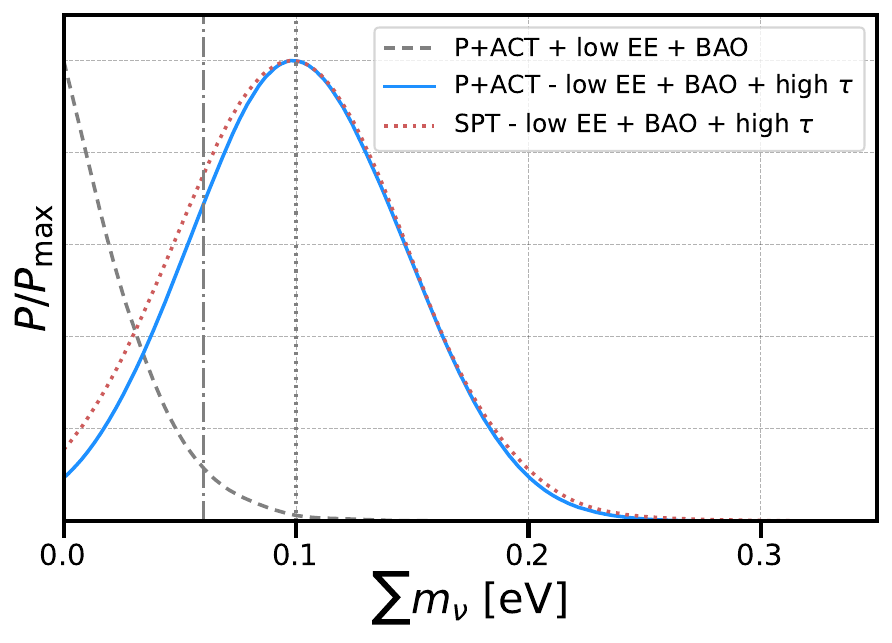}
    \caption{\textbf{Physical neutrino mass detection at $\mathbf{2\boldsymbol{\sigma}}$:}
    We find evidence for neutrino mass in the high-$\tau$ Universe at $2\sigma$ when varying the neutrino mass sum parameter $\sum  m_\nu$ (blue solid).
    For comparison, we also show the very tight constraint on $\sum  m_\nu$ from CMB+BAO data (gray dash-dotted) that favors zero neutrino mass.
    This evidence persists when using CMB data from SPT-3G DR1 (red dotted) rather than our fiducial ACT DR6 in combination with \textit{Planck}, indicating its robustness to choice of dataset and likelihood.
    We show gray vertical dash-dotted and dotted lines to indicate the minimal neutrino masses in the normal (NH) and inverted (IH) hierarchies.
    }
    \label{fig:mnu}
\end{figure}

The high-$\tau$ prior has several remarkable consequences (Fig.~\ref{fig:mnu}).
First, and most importantly, it converts the 
preference for ``negative'' neutrino mass into a tentative detection of a positive, physical mass: combining the CMB with DESI we infer $\sum m_\nu = 0.10^{+0.04}_{-0.05}$~eV (68\% C.L.), nominally the first cosmological detection of the neutrino mass at the $2\sigma$ level within $\Lambda$CDM, while the unphysical ``negative''-mass preference disappears entirely.
Second, the high-$\tau$ cosmology predicts low-redshift distances in excellent agreement with the DESI BAO measurements, thereby eliminating any need for dynamical dark energy to fit both datasets jointly; this conclusion remains valid, albeit slightly less significantly, when SNe data are included.
Third, it raises the CMB-inferred Hubble constant into better agreement with the DESI full-shape value, reducing the Hubble tension from $\approx 6\sigma$ to $4.4\sigma$ with the local Cepheid-calibrated SNe measurements -- a substantial reduction, though not a resolution, since a further increase in $\tau$ does not appreciably alleviate the tension; the resulting $H_0$ is moreover in no tension whatsoever 
with alternative local analyses~\cite{Freedman2025:CCHP,Wojtak2023:sne_multiple_pop}.

Methodologically, our analysis complements and extends previous high-$\tau$ proposals~\cite{Sailer2026:hightau,Giar2024:freetau_lowEE,Allali2026:reion_allali_H0_1,Jhaveri2025:hightau,Jhaveri2026:tau_features} in three ways:
First, we study the shifts of all the standard cosmological parameters as a function of $\tau$, paying special attention to $H_0$ and its CMB-independent determinations, including that from the DESI full-shape galaxy power spectrum and bispectrum analyses.
Second, building on~\cite{Green:2024xbb}, we investigate the lensing anomaly across multiple CMB likelihoods and its relation to the ``negative'' neutrino masses. Third, we compare the impact of uncalibrated supernovae -- including the updated DES Dovekie sample -- on the high-$\tau$ scenario.

Taken together, these results paint a tantalizingly consistent picture of our Universe across many independent datasets, built solely on the idea of the validity of the standard cosmological model. In the remainder of this work we demonstrate explicitly how a Universe more opaque than currently favored by large angular-scale \textit{Planck} polarization data can largely remove the need for new physics beyond $\Lambda$CDM on cosmological scales.
We describe the specific precision cosmological datasets 
that we use in Section~\ref{sec:data}, present our main results in Section~\ref{sec:results}, and comment on the significance of these results and the outlook in Section~\ref{sec:disc}.

\section{Data \label{sec:data}}
We use the following precision cosmology datasets to investigate the high-$\tau$ Universe:
\begin{itemize}
    \item \textbf{P+ACT + low EE}: We perform a baseline analysis with the $\texttt{SRoll2}$ \cite{2020A&A...635A..99P}\footnote{When including low-$\ell$ EE, we always use the $\texttt{SRoll2}$ likelihood.
    Given the modest shifts in, e.g., Ref.~\cite{Jhaveri2025:hightau}, considering an alternate low-$\ell$ EE likelihood for \textit{Planck} data is unlikely to affect our results.} low-$\ell$ \textit{EE} polarization likelihood included.
    For our fiducial high-$\ell$ \textit{TTTEEE} CMB dataset, we use the \textit{Planck} 2018 
    (PR3) \texttt{plik} + ACT DR6 high-$\ell$ \textit{TTTEEE} likelihoods where the \textit{Planck} data is cut at the low-$\ell$ multipoles where ACT does not have measurements \cite{actdr6_lcdm}. 
    We also use the low-$\ell$ TT Commander likelihood \cite{2020A&A...635A..99P}.
    We use the ACT DR6 + \textit{Planck} PR4 combined lensing likelihood \cite{Madhavacheril2024:act_lens,Qu2024:act_lens,Carron2022:pr4_lens}\footnote{We use the \texttt{actplanck\_baseline} variant.}. This is very similar to the \texttt{P-ACT-L} analysis of Ref.~\cite{actdr6_lcdm}.
    \item \textbf{P+ACT - low EE + high-$\tau$}:
    When considering the high-$\tau$ scenario, we use a $\tau = 0.11 \pm 0.006$ Gaussian prior and remove the low-$\ell$ EE polarization likelihood. 
    All other choices are the same as in ``P+ACT + low EE'',  above (using \textit{Planck} \texttt{plik}+ACT) unless otherwise specified. 
    \item \textbf{BAO}: We use the most recent DESI DR2 BAO distance measurements \cite{AbdulKarim2025:desi_dr2_bao}.
    \item \textbf{SNe}: For uncalibrated Type Ia SNe, we use the Pantheon+ \cite{Brout2022:pantheonplus} magnitude measurements as a function of redshift.
    We also consider those from the recently revised DES Dovekie sample \cite{Popovic2025:desdovekie}.
    \item \textbf{DESI-FS}: We consider the inference of $\tau$ from the joint analysis of CMB, BAO, and full-shape galaxy power spectrum and bispectrum data of DESI from Refs.~\cite{Chudaykin:2025aux,Ivanov2026:desi_dr1_4_kitchen_sink}. Here we consider the joint BAO + full-shape (3D + 2D) statistics, but use a Gaussian approximation to the cosmological parameters from the full-shape likelihood based on MCMC chains generated in Ref.~\cite{Ivanov2026:desi_dr1_4_kitchen_sink}. We have checked that this approximation is 
    highly accurate
    in combination with the CMB
    for the base $\Lambda$CDM model.
\end{itemize}

Since the combination of \textit{Planck} and ACT data is especially constraining, we follow Ref.~\cite{actdr6_lcdm} in using the $\texttt{plik}$ \textit{Planck} + ACT DR6 (compressed) likelihood as our fiducial likelihood.\footnote{See discussion of Fig.~37 of Ref.~\cite{actdr6_lcdm} for the ACT collaboration's justification of this choice as well as Ref.~\cite{Jense2026:planck_cmb_lhood}.}
However, to investigate the dependence of our results on this choice, we also consider several variations of primary CMB and CMB lensing likelihoods. 
In particular:
\begin{itemize}
    \item \textbf{CS(PR4)+ACT}: As a \textit{Planck}-internal check of the high-$\ell$ likelihood dependence and CMB map, we consider the CamSpec \textit{TTTEEE} high-$\ell$ likelihood \cite{Rosenberg2022:pr4_camspec} applied to PR4 (NPIPE; \cite{Planck_npipe}) maps when combined with higher $\ell$ measurements from ACT DR6~\cite{actdr6_lcdm} where we also add the ACT DR6 + \textit{Planck} PR4 lensing data. 
    \item \textbf{W+ACT}: We form another \textit{Planck}-independent ACT data combination by considering WMAP9 \cite{2013ApJS..208...20B} data as the lower multipoles are inaccessible to ACT, following the ``W-ACT'' analysis of Ref.~\cite{actdr6_lcdm}, but where we also add the ACT DR6 + \textit{Planck} PR4 lensing data\footnote{We purposely refer to this data combination with ``$+$'' to distinguish it from ``W-ACT'' of Ref.~\cite{actdr6_lcdm}, which does not include lensing data.}.
    \item \textbf{SPT}: Independent of both \textit{Planck} and ACT, we also consider SPT-3G primary and lensing data \cite{GeMillea2025:spt_3g_polzn}.
    We use the \texttt{candl} \cite{Balkenhol:2024sbv} foreground-marginalized ``lite'' high-$\ell$ likelihood for $TTTEEE$ (``SPT-3G D1 T\&E'') as presented in Ref.~\cite{Camphuis2025:SPT3GD1}.
    We use the lensing likelihood (EE only) from Ref.~\cite{GeMillea2025:spt_3g_polzn} as implemented in the \texttt{act\_dr6\_spt\_lenslike} Cobaya likelihood with the spt3g variant\footnote{This likelihood is implemented in the \href{https://github.com/qujia7/spt_act_likelihood}{\texttt{spt\_act\_likelihood}} repository as the ``spt3g'' variant. Despite the ``ACT'' in the name, it only uses SPT-3G data.}.
\end{itemize}

When running Markov-Chain Monte Carlo (MCMC) chains in \texttt{Cobaya} \cite{cobaya},
runs are continued until Gelman-Rubin convergence of $\hat{R}-1 <0.01$ among 4 or 8 chains, with the exception of the alternative likelihood chains (involving CS(PR4)+ACT, W+ACT), which are run to $\hat{R}-1 <0.02$. 

\section{Results \label{sec:results}}
In this Section, we begin by presenting parameter constraints from the CMB primary anisotropies and lensing in Sec.~\ref{subsec:cmb}: we first show how the high-$\tau$ prior shifts the $\Lambda$CDM parameters inferred from the CMB alone in Sec.~\ref{subsubsec:lcdm_cmb}, and then examine the \textit{Planck} lensing anomaly in the high-$\tau$ Universe -- its dependence on the choice of likelihood and its relation to the preference for ``negative'' neutrino mass in Sec.~\ref{subsec:lens_neg_nu_mass}. 
We then consider the concording power of our high-$\tau$ Universe in the context of DESI DR2 BAO distance measurements combined with those of the CMB in Sec.~\ref{subsec:bao_distance}. There we find evidence for a physical (positive) neutrino mass in Sec.~\ref{subsec:nu_mass} and a lack of evidence for evolving dark energy, both of which largely persist once uncalibrated Type Ia supernovae (SNe) are included.
Next, we assess the extent to which precision $\Lambda$CDM CMB and large-scale structure data permit a high value of $\tau$ as a solution to the local Hubble tension in Sec.~\ref{subsubsec:highh0tau}.
Finally, we provide a CMB-independent consistency inference of $\tau$ from the DESI full-shape data, which requires no low-$\ell$ EE polarization information in Sec.~\ref{subsec:fullshape}.

\subsection{CMB Primary Anisotropies \& Lensing \label{subsec:cmb}}

\subsubsection{$\Lambda$CDM parameters \label{subsubsec:lcdm_cmb}}

\begin{figure*}
    \centering
    \includegraphics[width=0.95\linewidth]{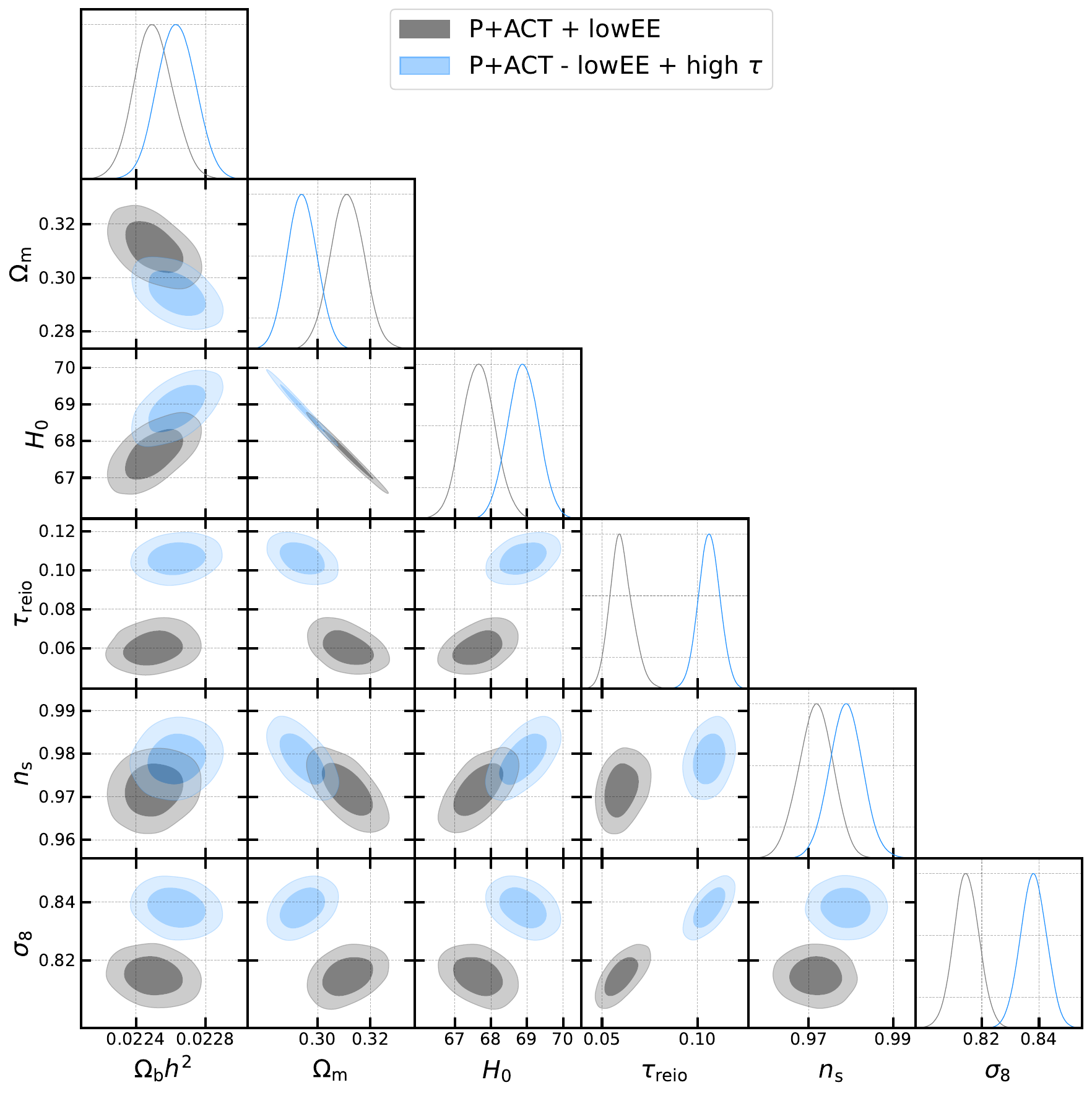}
    \caption{\textbf{$\Lambda$CDM parameters in a high-$\tau$ Universe:} CMB-only constraints on the $\Lambda$CDM parameters. 
    The high-$\tau$ results are shown in blue for our fiducial \textit{Planck}+ACT combination without the low-$\ell$ \textit{Planck} EE polarization data.
    For comparison we show the fiducial constraints when the low-$\ell$ EE likelihood is included in gray.
    The $\Lambda$CDM parameters are largely consistent, but, relative to the case with low-$\ell$ EE polarization (and so low $\tau$), $H_0$ and $\sigma_8$ each shift up by $\sim2\sigma$, while $\Omega_m$ shifts down by $\sim2\sigma$.
    }
    \label{fig:lcdm_triangle}
\end{figure*}

\begin{table*}[!htb]
    \centering
    \renewcommand{\arraystretch}{1.2}
    \setlength{\tabcolsep}{5pt}
    \begin{tabular}{@{}l ccccc@{}}
        \hline\hline
        \textbf{Dataset}
        & $\omega_{\mathrm{m}}$
        & $H_0$
        & $\Omega_{\mathrm{m}}$
        & $\sigma_8$
        & $S_8$ \\
        \hline
        \multicolumn{6}{@{}l}{\textbf{$\Lambda$CDM}} \\
        \quad P+ACT $+$ low-$\ell$ EE
            & $0.1424 \pm 0.0010$ & $67.65_{-0.46}^{+0.44}$ & $0.311 \pm 0.006$ & $0.814 \pm 0.004$ & $0.830 \pm 0.011$ \\
        \quad P+ACT $-$ low-$\ell$ EE $+$ h$\tau$
            & $0.1396 \pm 0.0010$ & $68.89\pm0.43$ & $0.294^{+0.005}_{-0.006}$ & $0.838 \pm 0.005$ & $0.830_{-0.010}^{+0.011}$ \\
        \quad P+ACT $+$ low-$\ell$ EE $+$ BAO
            & $0.1407 \pm 0.0006$ & $68.43 \pm 0.27$ & $0.300 \pm 0.004$ & $0.813 \pm 0.005$ & $0.813 \pm 0.007$ \\
        \quad P+ACT $-$ low-$\ell$ EE $+$ h$\tau$ $+$ BAO
            & $0.1397 \pm 0.0006$ & $68.85^{+0.26}_{-0.27}$ & $0.295 \pm 0.003$ & $0.838 \pm 0.004$ & $0.831 \pm 0.007$ \\
        \addlinespace
        \multicolumn{6}{@{}l}{\textbf{$\Lambda$CDM $+ \sum m_\nu$}} \\
        \quad P+ACT $-$ low-$\ell$ EE $+$ h$\tau$ $+$ BAO
            & $0.1398 \pm 0.006$ & $68.61 \pm 0.36$ & $0.297 \pm 0.004$ & $0.829 \pm 0.011$ & $0.824 \pm 0.010$ \\
        \addlinespace
        \multicolumn{6}{@{}l}{\textbf{$w_0 w_a$CDM}} \\
        \quad P+ACT $-$ low-$\ell$ EE $+$ h$\tau$ $+$ BAO
            & $0.1398 \pm 0.0008$ & $66.17^{+1.88}_{-1.87}$ & $0.320^{+0.017}_{-0.020}$ & $0.816 \pm 0.016$ & $0.842^{+0.012}_{-0.013}$ \\
        \quad P+ACT $-$ low-$\ell$ EE $+$ h$\tau$ $+$ BAO $+$ SNe
            & $0.1396 \pm 0.0008$ & $67.48 \pm 0.59$ & $0.307 \pm 0.006$ & $0.826 \pm 0.008$ & $0.835 \pm 0.008$ \\
        \hline
    \end{tabular}
    \caption{%
        \textbf{Parameter concordance with high $\tau$:}
        $\Lambda$CDM parameter constraints under the ``high-$\tau$'' (h$\tau$) scenario ($\tau = 0.11 \pm 0.006$) for the (physical) matter overdensity parameter $\Omega_{\mathrm{m}}$ ($\omega_{\mathrm{m}}$), the Hubble parameter $H_0$, and the amplitude of matter density fluctuations $\sigma_8$ for several cosmological models (standard $\Lambda$CDM, $\Lambda$CDM\,$+\sum m_\nu$ with free neutrino mass, and evolving dark energy $w_0 w_a$CDM).
        For reference, we also consider the $\Lambda$CDM constraints from the CMB alone with low-$\ell$ EE polarization data included, which favors $\tau = 0.06 \pm 0.006$.
        The high-$\tau$ scenario leads to: (1) increased $H_0$, mitigating the Hubble tension; (2) decreased $\Omega_{\mathrm{m}}$ as inferred from the CMB, in better agreement with the BAO constraints; and (3) a matter clustering amplitude $S_8$ that does not change significantly from the case where \textit{Planck} low-$\ell$ EE polarization data is included.
        Data combinations are as described in Section~\ref{sec:data}.
        }
    \label{tab:main}
\end{table*}

The use of the high-$\tau$ prior produces significant shifts in several of the CMB-inferred parameters.
Results for our fiducial CMB dataset (P+ACT) are displayed in
Fig.~\ref{fig:lcdm_triangle} and Table~\ref{tab:main}.
We show the high-$\tau$ model for both our fiducial \textit{Planck}+ACT CMB data combination (blue contours, including ACT lensing)\footnote{When combined with the less-constraining SPT data (including SPT-only lensing), we find qualitatively similar results, see Appendix~\ref{subsec:spt}}.
For comparison, we also show the baseline CMB results based on the usual inclusion of the low-$\ell$ EE likelihood ($\texttt{SRoll2}$, black contours). 
After removing the low-$\ell$ EE contribution, we find a $\Delta\chi^2 = \chi^2_{\mathrm{h}.\tau.}-\chi_{\mathrm{(P+ACT+low~EE)-low~EE}}^2=-1.9$.
In particular, when applying the high-$\tau$ prior, we see significant shifts up (at more than $2\sigma$) of the matter fluctuation amplitude $\sigma_8$ and the Hubble parameter $H_0$, while the matter overdensity parameter $\Omega_m$ (or $\omega_m$) shifts down by $3\sigma$.
The changes in the $\omega_b$ and $n_s$ posteriors are marginal, as they are consistent between the baseline results and the high-$\tau$ case at 1$\sigma$.
We also notice that the lensing amplitude parameter $S_8$ is unchanged in the high-$\tau$ case (see Tab.~\ref{tab:main}).

These shifts can be straightforwardly understood in the following way.
An increase in $\tau$ (at high significance) leads to significantly higher values of $A_s$ (and so $\sigma_8$) as preferred by the high-$\ell$ primary CMB through the $A_s-\tau$ degeneracy (via the inferred amplitude $A_s e^{-2\tau}$).
For a fixed, well-measured, lensing amplitude $S_8$, this implies a smaller value of $\Omega_m$, which is free to shift down only to the extent that this shift can be compensated by a rise in $H_0$ through the CMB $\Omega_m h^3$ degeneracy.
Indeed, these are exactly the shifts seen in Fig.~\ref{fig:lcdm_triangle}, where the $\Omega_m-H_0$ contour shifts to a lower-$\Omega_m$ region of the CMB degeneracy, while $\sigma_8$ shifts up.

The shifts of $\Omega_m$ and $H_0$ along the CMB degeneracy produce better agreement with low-redshift $\Lambda$CDM constraints from DESI.
The higher value of $H_0=68.89\pm 0.42$ km/s/Mpc (see the second row of Tab.~\ref{tab:main}) agrees well with the value from DESI+BBN of $H_0=68.51\pm 0.58$ km/s/Mpc \cite{AbdulKarim2025:desi_dr2_bao}.
It also agrees perfectly with the DESI-FS value $H_0=69.09\pm 0.37$ km/s/Mpc from the DESI reanalysis of Ref.~\cite{Ivanov2026:desi_dr1_4_kitchen_sink}. 
This agreement suggests that our extrapolation based on the Gaussian likelihood approximation for the $\Lambda$CDM parameters is accurate enough for the purposes of this exploratory study.
We discuss the limits of this extrapolation further in Section~\ref{subsubsec:highh0tau}.\footnote{We might anticipate that, due to the strong degeneracy between $H_0$ and $N_\mathrm{eff}$, the higher value of $\tau$ considered here may shift upward the low value of $N_\mathrm{eff}$ measured by ACT \cite{act_exts_calabrese}. 
A similar scenario has been explored when removing low-$\ell$ EE data entirely \cite{Goldstein2026:neff}; there the upward shift of $N_\mathrm{eff}$ was $<0.5\sigma$ when combining \textit{Planck}, ACT, and SPT data.}

The lower 
CMB value of $\Omega_m$ in the high-$\tau$ Universe also approaches the value preferred by the DESI DR2 BAO data, as the CMB-based BAO distance ratios predicted by the CMB are low enough to agree with the data (see Section~\ref{subsec:bao_distance} for a discussion).
In addition, our CMB-based $\Omega_m= 0.294_{-0.006}^{+0.005}$
agrees well with the CMB-independent $\Omega_m=0.296\pm0.007$ inference 
from the full-shape of the DESI galaxy power spectra 
and bispectra~\cite{Chudaykin:2025aux,Ivanov2026:desi_dr1_4_kitchen_sink}.
This is an important observation, as the full-shape  
$\Omega_m$ measurement stems from the shape of the transfer functions~\cite{Ivanov:2019pdj}, which depend on $\omega_m$ in $\Lambda$CDM. In principle, 
this is an ``early-Universe''
measurement, which should be contrasted with the BAO-based 
``late-Universe''
$\Omega_m$ measurement
driven by the Alcock-Paczy\'nski
effect. In physical models 
designed to address the 
``phantom behavior'' of DESI's evolved dark energy, 
such as KMIX~\cite{Toomey:2025yuy} or Dark QCD~\cite{Khoury:2025txd}, 
the values of the 
effective 
``early'' and ``late''
matter abundances\footnote{The effective 
``early''
value in this context is the 
matter abundance at
recombination extrapolated to $z=0$ using the 
standard Friedmann equation. } 
are different.
The fact that full-shape, BAO, and the high-$\tau$ CMB data produce compatible values of $\Omega_m$
is a non-trivial 
consistency test of 
the $\Lambda$CDM model across
different datasets.

\begin{figure}
    \centering
    \includegraphics[width=0.95\linewidth]{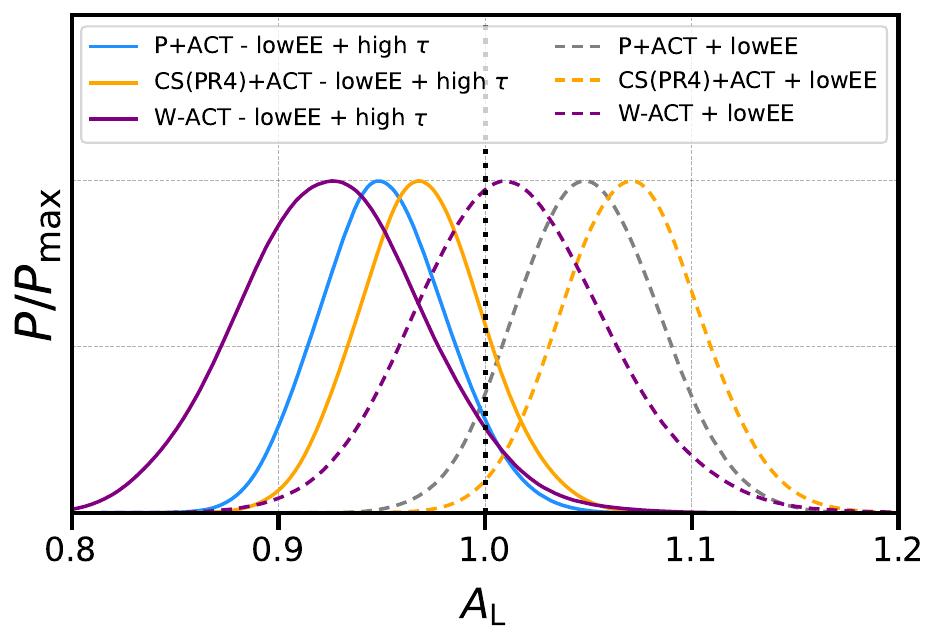}\hfill
    \caption{\textbf{Reversing the lensing anomaly:}
    We illustrate the removal of the low-$\ell$ $EE$ polarization CMB data and adding our high-$\tau$ prior when going from dashed to solid lines for different CMB likelihoods: PR3 (\texttt{plik}) + ACT (gray), CS(PR4) + ACT (orange) and W + ACT (purple).
    The baseline CMB results with low-$\ell$ $EE$ polarization for our fiducial $\texttt{plik}$ likelihood (gray dashed) exhibit a $\sim2\sigma$ preference for large values of $A_\mathrm{L}$.
    Also for CamSpec+ACT (orange dashed), the likelihood has a more significant preference for $A_\mathrm{L}>1$ while the W+ACT data demonstrates weak/no preference for $A_\mathrm{L}$ (purple dashed).
    When $\tau$ is high, there is a similarly significant  preference for low $A_\mathrm{L}$ across all likelihoods: somewhat for \texttt{plik} (blue solid) and W+ACT (purple solid), and mildly for  for CamSpec+ACT (orange solid).
    This reversal reflects the preference for high CMB lensing power amplitude relative to that predicted by the primary CMB in the high-$\tau$ Universe.
    }
    \label{fig:alens_1d}
\end{figure}

\subsubsection{Lensing anomaly and ``negative'' neutrino mass \label{subsec:lens_neg_nu_mass}}
\begin{table*}[!htb]
    \centering
    \renewcommand{\arraystretch}{1.2}
    \setlength{\tabcolsep}{5pt}
    \begin{tabular}{@{}l ccccc@{}}
        \hline\hline
        \textbf{Dataset}
        & $\omega_{\mathrm{m}}$
        & $H_0$
        & $\Omega_{\mathrm{m}}$
        & $\sigma_8$
        & $S_8$ \\
        \hline
        \multicolumn{6}{@{}l}{\textbf{$\Lambda$CDM}} \\
        \quad CS(PR4)+ACT $+$ low-$\ell$ EE
            & $0.1415 \pm 0.0010$ & $67.95 \pm 0.43$ & $0.307 \pm 0.006$ & $0.813 \pm 0.005$ & $0.822 \pm 0.010$ \\
        \quad W+ACT $+$ low-$\ell$ EE
            & $0.1434 \pm 0.0013$ & $67.51^{+0.52}_{-0.50}$ & $0.315 \pm 0.007$ & $0.816 \pm 0.005$ & $0.835 \pm 0.012$ \\
        \quad CS(PR4)+ACT $+$ low-$\ell$ EE $+$ h$\tau$
            & $0.1392 \pm 0.0009$ & $68.99 \pm 0.41$ & $0.292 \pm 0.005$ & $0.836 \pm 0.004$ & $0.826 \pm 0.010$ \\
        \quad W+ACT $+$ low-$\ell$ EE $+$ h$\tau$
            & $0.1387^{+0.0011}_{-0.0012}$ & $69.42^{+0.47}_{-0.46}$ & $0.288 \pm 0.006$ & $0.835 \pm 0.005$ & $0.818 \pm 0.011$ \\
        \hline
        \multicolumn{6}{@{}l}{\textbf{$A_{\mathrm{L}}$}} \\
        \quad CS(PR4)+ACT $+$ low-$\ell$ EE
            & $0.1401 \pm 0.0011$ & $68.59 \pm 0.52$ & $0.298 \pm 0.007$ & $0.801 \pm 0.007$ & $0.798 \pm 0.014$ \\
        \quad CS(PR4)+ACT $-$ low-$\ell$ EE $+$ h$\tau$
            & $0.1399 \pm 0.0011$ & $68.67^{+0.52}_{-0.51}$ & $0.297 \pm 0.007$ & $0.842 \pm 0.007$ & $0.838 \pm 0.015$ \\
        \quad W+ACT $+$ low-$\ell$ EE
            & $0.1432 \pm 0.0024$ & $67.56 \pm 0.97$ & $0.314^{+0.013}_{-0.015}$ & $0.814 \pm 0.011$ & $0.833^{+0.028}_{-0.030}$ \\
        \quad W+ACT $-$ low-$\ell$ EE $+$ h$\tau$
            & $0.1422 \pm 0.0024$ & $67.93^{+0.99}_{-0.98}$ & $0.309^{+0.013}_{-0.015}$ & $0.854 \pm 0.012$ & $0.867 \pm 0.031$ \\
        \hline
        \multicolumn{6}{@{}l}{\textbf{$\sum m_{\nu,\mathrm{eff}}$}} \\
        \quad CS(PR4)+ACT $+$ low-$\ell$ EE
        & $0.1374_{-0.0022}^{+0.0040}$
        & $70.28_{-1.98}^{+1.31}$
        & $0.279_{-0.016}^{+0.023}$
        & $0.850_{-0.033}^{+0.022}$
        & $0.818 \pm 0.010$
        \\
        \quad CS(PR4)+ACT $-$ low-$\ell$ EE $+$ h$\tau$
        & $0.1418_{-0.0021}^{+0.0024}$
        & $67.08_{-1.71}^{+1.40}$
        & $0.316_{-0.020}^{+0.019}$
        & $0.805_{-0.027}^{+0.023}$
        & $0.825 \pm 0.010$
        \\
        \quad W+ACT $+$ low-$\ell$ EE
        & $0.1451_{-0.0045}^{+0.0047}$
        & $66.43_{-2.60}^{+2.65}$
        & $0.331_{-0.040}^{+0.033}$
        & $0.800_{-0.035}^{+0.033}$
        & $0.838_{-0.014}^{+0.016}$
        \\
        \quad W+ACT $-$ low-$\ell$ EE $+$ h$\tau$
        & $0.1463 \pm 0.0034$
        & $64.62_{-2.19}^{+1.88}$
        & $0.352 \pm 0.030$
        & $0.769_{-0.031}^{+0.028}$
        & $0.831 \pm 0.013$
        \\
        \quad CS(PR4)+ACT $+$ low-$\ell$ EE $+$ BAO
            & $0.1400 \pm 0.0007$ & $68.94^{+0.35}_{-0.39}$ & $0.295 \pm 0.004$ & $0.831^{+0.010}_{-0.012}$ & $0.823^{+0.009}_{-0.010}$ \\
        \quad CS(PR4)+ACT $-$ low-$\ell$ EE $+$ h$\tau$ $+$ BAO
            & $0.1396 \pm 0.0006$ & $68.61^{+0.39}_{-0.38}$ & $0.297 \pm 0.004$ & $0.828^{+0.011}_{-0.012}$ & $0.823 \pm 0.010$ \\
        \quad W+ACT $+$ low-$\ell$ EE $+$ BAO
            & $0.1407 \pm 0.0007$ & $68.97^{+0.36}_{-0.40}$ & $0.296^{+0.004}_{-0.005}$ & $0.832^{+0.010}_{-0.013}$ & $0.827 \pm 0.010$ \\
        \quad W+ACT $-$ low-$\ell$ EE $+$ h$\tau$ $+$ BAO
            & $0.1397 \pm 0.0007$ & $68.64^{+0.39}_{-0.38}$ & $0.297 \pm 0.004$ & $0.822^{+0.011}_{-0.012}$ & $0.818 \pm 0.010$ \\
        \hline
    \end{tabular}
    \caption{%
        \textbf{Alternative CMB likelihoods ($A_{\mathrm{L}}$ \& $\sum m_{\nu,\mathrm{eff}}$):}
        Similar to Tab.~\ref{tab:main}, but for the alternative likelihoods presented in Section.~\ref{subsec:lens_neg_nu_mass}.
        Here we consider the combined CamSpec + ACT likelihood with the \textit{Planck} PR4 maps (CS(PR4)+ACT) and the combined WMAP9 + ACT likelihood, both with combined CMB lensing from \textit{Planck}+ACT (see Section~\ref{sec:data}).
        We show results for $\Lambda$CDM as well as the single-parameter lensing anomaly ($A_{\mathrm{L}}$) and effective neutrino mass ($\sum m_{\nu,\mathrm{eff}}$) extended models. \
        }
    \label{tab:al_mnu_alt_like}
\end{table*}

\paragraph{Lensing anomaly:}
Large-scale structure generates coherent deflections of CMB photons through weak lensing.
CMB data is sensitive to this effect at the 2-point level (the smoothing of oscillations in the CMB scalar anisotropy spectra) and at the lensing reconstruction\footnote{As a shorthand, we will refer to this as ``4-point'' information following the standard quadratic estimator reconstruction, but note that Bayesian methods are also used to reconstruct the lensing potential \cite[e.g.,][]{hirataseljak,Millea2020:delens_bayes,GeMillea2025:spt_3g_polzn}. } level (reconstruction of the LSS lensing potential power $C_L^{\phi\phi}$ via the trispectrum).
Within the $\Lambda$CDM model, the amplitude of the power spectrum of this lensing signal is fully specified by the model parameters, such as $A_s$, $\Omega_m$.
To test for the consistency of the $\Lambda$CDM parameters preferred by the primary CMB and those preferred by the lensing signal (from both the 2-point and 4-point, or reconstruction, information) one can formally introduce a parameter
$A_\mathrm{L}$ that rescales the CMB lensing
power spectrum, $C_L^{\phi\phi}\to A_\mathrm{L} C_L^{\phi\phi}$, and which 
alters both the 2-point and 4-point effects \cite{Calabrese2008:alens,planck13,planck15}.
Thus, the $A_\mathrm{L}$ rescaling allows one to quantify
to what extent the actual lensing signal 
is consistent with the predictions
based on the fit to the CMB primary anisotropies.

\begin{figure*}
    \centering
    \includegraphics[width=0.49\linewidth]{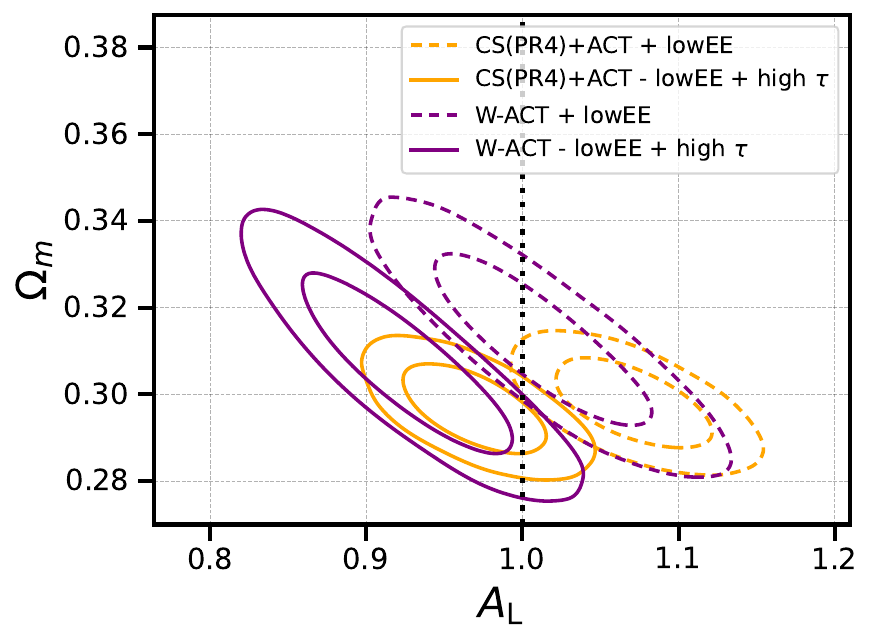}
    \includegraphics[width=0.49\linewidth]{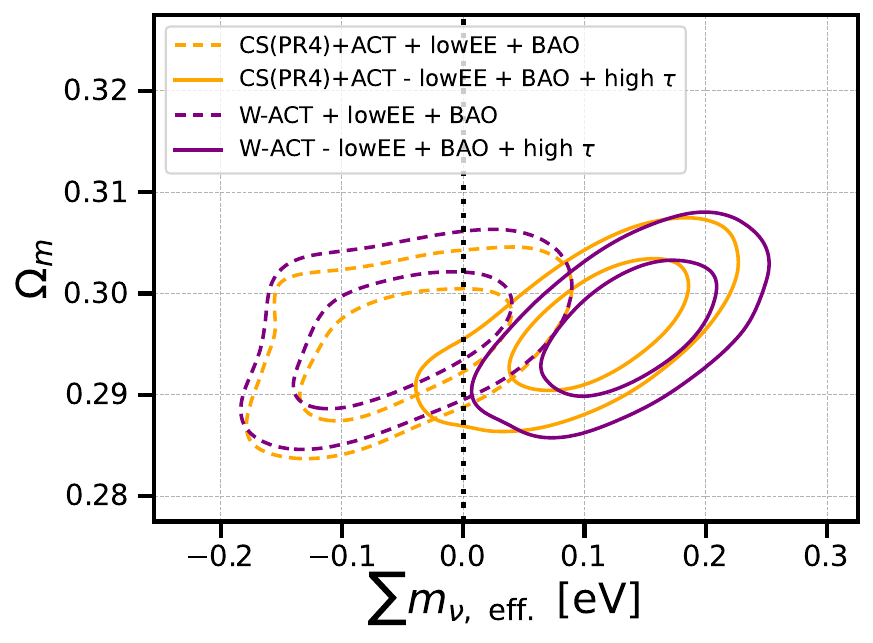}
    \caption{\textbf{ $\Omega_m$, $A_\mathrm{L}$, and $\sum m_{\nu,\mathrm{eff}}$ in
    alternative CMB likelihoods:}
    \textit{Left}: 
    The baseline ``+ low EE'' model (dashed) is shown for the CamSpec+ACT and W+ACT CMB data combinations in the $\Omega_m-A_\mathrm{L}$ plane. 
    Solid curves show the high-$\tau$ model results.
    The low EE baseline result favors a positive lensing anomaly ($A_\mathrm{L}>1$) for CamSpec+ACT, but not for W+ACT, while the high-$\tau$ model removes the lensing anomaly in CamSpec+ACT, while introducing a preference for $A_\mathrm{L}<1$ in W+ACT.
    Conditioned on no lensing anomaly ($A_\mathrm{L}=1$), the low EE baseline prefers higher $\Omega_m$ (and lower $H_0$ and $\sigma_8$), while the reverse is true in the high-$\tau$ model.
    \textit{Right}: 
    In a ``negative'' effective neutrino mass model, the high-$\tau$ models prefers positive neutrino mass \textit{regardless} of the CMB likelihood used when combined with DESI DR2 BAO data.
    This suggests that the preference for negative neutrino mass is \textit{not} simply explained as the preference for $A_\mathrm{L}$, which differs significantly between likelihoods.
    }
    \label{fig:AL_mnu_Omm}
\end{figure*}

Previous analyses of 
the \textit{Planck} 2018
PR3
\texttt{plik} likelihood (\textit{Planck} 2018) have found this scaling 
parameter to deviate from its
consistency value $A_\mathrm{L}=1$ at the level of $\approx (2-3)\sigma$~\cite{planck_lens_18}.
This preference for 
more lensing than 
predicted by the \textit{Planck} 2018 CMB primaries
within the base $\Lambda$CDM model
is called ``the lensing anomaly.''
In particular, analyses of the primary \textit{Planck} 2018 data have found a nearly $3\sigma$ discrepancy with the base $\Lambda$CDM predictions,
$A_\mathrm{L} = 1.180 \pm 0.065$.
This value drops to $A_\mathrm{L}=1.071^{+0.038}_{-0.042}$ upon addition of the 
lensing reconstruction data.\footnote{$A_\mathrm{L}$ has also been explored in combination with other cosmological datasets more recently, e.g. in Refs.~\cite{Addison2016:discordance,RoyChoudhury:2024free_everything,Garcia-Quintero2025:spt_baodr2}.}
The lensing anomaly in the 
\textit{Planck} data and its resolution are driven largely by two effects: differences in the lower- and higher-$\ell$ $TT$ likelihoods and the low-$\ell$ EE polarization likelihoods.
Recent alternative likelihoods
have found improved consistency 
in terms of $A_\mathrm{L}$. 
In particular, the
low-$\ell$ LoLLiPoP and high-$\ell$ HiLLiPoP likelihoods based on the 
newer PR4 \textit{Planck} maps have yielded $A_\mathrm{L}$ consistent with 1 within $\approx 2\sigma$~\cite{Tristram2024:pr4}, and
Ref.~\cite{Rosenberg2022:pr4_camspec} also finds $A_\mathrm{L} = 1.095\pm0.056$ using the CamSpec PR4 TTTEEE likelihood. 
In addition, the combination of 
WMAP and ACT data (W-ACT) has yielded
$A_\mathrm{L}=1.043 \pm 0.049$, consistent with unity
within $1\sigma$ \cite{actdr6_lcdm}.

From this history, it is clear that different CMB likelihoods support different levels of preference for the lensing anomaly $A_\mathrm{L}>1$.
As a result we consider several of these likelihoods in what follows, and first describe the preference for the lensing anomaly in the most closely related likelihoods and how the inferred $\Lambda$CDM parameters shift, before discussing how they change in the high-$\tau$ Universe.

In the \textit{Planck} 2018 analysis, the shape of the CMB primaries put tight constraints on 
$\omega_m$ and $\Omega_m$,
while the low-$\ell$ EE data fix $\tau$ and hence $A_s$ through their best-measured combination $A_se^{-2\tau}$. 
In this configuration the fit does not have enough freedom to produce 
a large lensing signal, resulting in evidence 
for $A_\mathrm{L}>1$. 
For W-ACT this 
shift is sufficient to 
account for the lensing excess, making 
the base $\Lambda$CDM cosmology consistent with $A_\mathrm{L}=1$ \cite{actdr6_lcdm}.
The $\Lambda$CDM lensing amplitude is 
approximately 
set by 
$\sigma_8\Omega_m^{0.25}$,
which has a large 
$\approx 80\%$ correlation
with $\Omega_m$
in the primary CMB data 
(without the lensing four-point function).\footnote{This correlation coefficient is extracted from the base \textit{Planck} 2018 MCMC chains without CMB lensing \cite{planck18}} 
Once the low-$\ell$ EE data 
has fixed $A_s$ through $\tau$, $\Omega_m$
remains the only available
channel to account for the lensing excess. 
In the CamSpec PR4+ACT (CS(PR4)+ACT) likelihood, the $\Omega_m$ maximum a posteriori (MAP) value is somewhat low, which cannot account for the lensing excess, 
and hence some 
$2\sigma$ evidence for $A_\mathrm{L}>1$ remains \cite{Rosenberg2022:pr4_camspec}. 
However, the 
W-ACT 
data allows for a large value of $\Omega_m =  0.326 \pm 0.011$,
which naturally predicts 
more lensing than \textit{Planck} 2018, and hence 
removes the ``lensing anomaly'' (see Tab.~5, eqn.~45,
of Ref.~\cite{actdr6_lcdm}\footnote{Note that these values are quoted for the W-ACT combination which does not include lensing 4-point/reconstruction data. For completeness, we quote the value including lensing data here, $A_\mathrm{L} = 1.018_{-0.045}^{+0.041}$, obtained from the publicly available \href{https://lambda.gsfc.nasa.gov/product/act/act_dr6.02/act_dr6.02_chains_prod_table.html}{ACT DR6 chains}.}).

We find that the W+ACT + low EE likelihood has no change in the central values of its parameters when $A_\mathrm{L}$ is varied; the constraints only weaken (see Table~\ref{tab:al_mnu_alt_like}).
On the other hand, for CS(PR4)+ACT + low EE, the value of $\Omega_m$ drops $1\sigma$ and the value of $\sigma_8$ drops almost $2\sigma$ (here $\omega_m$ drops more than $1\sigma$ because we are moving along the $\Omega_m h^3$ degeneracy; $H_0$ increases by more than $1\sigma$) when $A_\mathrm{L}$ is varied.
In $\Lambda$CDM, CS(PR4)+ACT + low EE and W+ACT + low EE disagree with each other at $>1\sigma$ on the value of $\Omega_m$ (and $\omega_m$), and so also on the value of $S_8$.
The presence of significant $A_\mathrm{L}>1$ for CS(PR4)+ACT + low EE allows the parameters to stay the same while matching the measured lensing amplitude.
However, this does not resolve the underlying $\gtrsim1\sigma$ mismatch of the $\Lambda$CDM parameters, which \textit{can} be solved by raising $\tau$, as the values shift toward lower $\Omega_m$, higher $H_0$, and higher $\sigma_8$.

Let us now investigate the 
lensing excess in the context 
of our high-$\tau$ Universe, as it is known that the value of $\tau$ is connected to the amplitude of $A_\mathrm{L}$ \cite[e.g.][]{Giar2024:freetau_lowEE,Jhaveri2025:hightau,LoverdeWeiner2024:BAO,Green:2024xbb}.
The results of these analyses are shown in Fig.~\ref{fig:alens_1d}.
There we see that the high-$\tau$
Universe produces 
$A_\mathrm{L}$ consistent with one within
$2\sigma$ in all analyses. 
In particular, 
we find 
\begin{equation}
    A_\mathrm{L} = 0.95 \pm 0.03
    \label{eqn:cmb_alens_const}
\end{equation}
from the full combination
of CMB data 
including 
\textit{Planck} 2018 and ACT.
Similarly, we find $A_\mathrm{L}$ consistent with $A_\mathrm{L}=1$
in the W+ACT high-$\tau$
analysis, though the central value 
is $\approx 2\sigma$
lower than unity. 
In this case 
we have a 
lensing deficit 
instead of an excess,
i.e. the high-$\tau$ Universe ``reverses''
the lensing anomaly.
We stress, however, that the preference for 
low $A_\mathrm{L}$ in our high-$\tau$ analysis is 
conceptually different 
from the preference for large $A_\mathrm{L}$
in the standard analyses.
Our preference for $A_\mathrm{L}<1$ 
can be interpreted as 
a result of assuming 
the lowest possible neutrino mass in the
base $\Lambda$CDM model. 
Once the neutrino masses are allowed to vary, the data prefers a value larger than $0.06$ eV, restoring the consistency with $A_\mathrm{L}= 1$. This ``reverse lensing anomaly'' is  
distinct from the original
``lensing anomaly'' (with 
$A_\mathrm{L}>1$) because the latter case requires quite drastic 
beyond-$\Lambda$CDM 
scenarios~\cite[e.g.][]{Craig2024:neg_nu}, while the former is fully consistent with physical neutrinos.

Our high-$\tau$ Universe exhibits more concordance between low-$\ell$ and high-$\ell$ data
than the standard \textit{Planck} 2018 baseline in terms of the amplitude of the lensing power spectrum. We do find 
a tentative $\approx 2\sigma$ preference for $A_\mathrm{L}<1$, but as we shall see, this 
is a consequence
of total 
neutrino mass being 
slightly 
larger than the minimal 
value $0.06$~eV.

\begin{figure}
    \centering
    \includegraphics[width=0.95\linewidth]{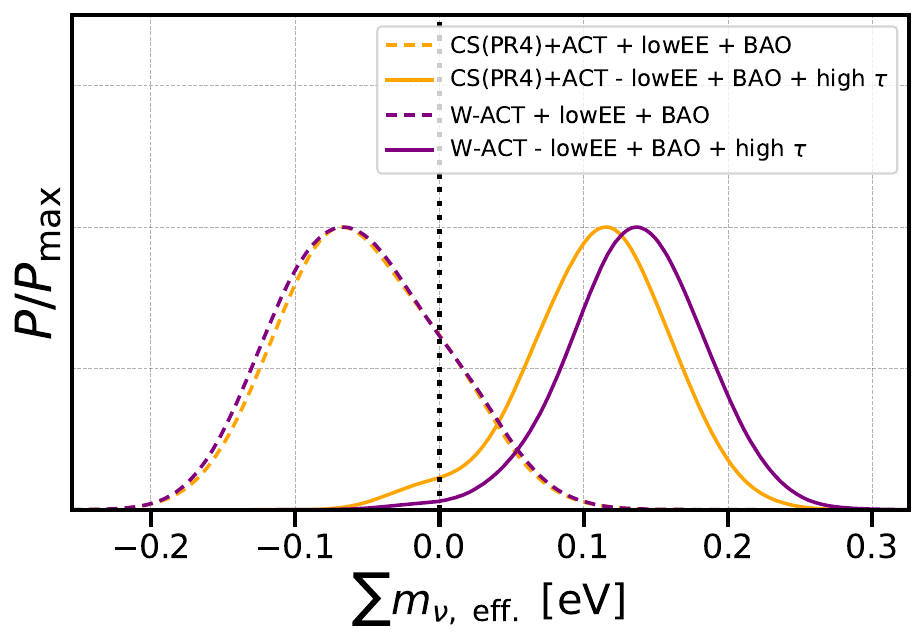}
    \caption{\textbf{No ``negative'' neutrino mass:}
    We show that when an effective ``negative'' neutrino mass $\sum m_{\nu,\mathrm{eff.}}$ is permitted as a device to fit the reconstructed CMB lensing power spectrum, it is disfavored by CMB+BAO data in the high-$\tau$ Universe.
    We show a combined CamSpec+ACT likelihood with the PR4 maps (orange), as well as the W+ACT likelihood (purple).
    Dashed curves show the results with the \textit{Planck} $\texttt{SRoll2}$ low-$\ell$ $EE$ likelihood, while solid curves show the predictions for the high-$\tau$ case.
    The rejection of negative neutrino mass is significant \textit{irrespective} of the likelihood used, despite their differing levels of preference for $A_\mathrm{L}\neq1$ in Fig.~\ref{fig:alens_1d}
    }
    \label{fig:negnumass_1d}
\end{figure}

\paragraph{``Negative'' neutrino mass.}
The lensing anomaly is 
tightly connected with another tension
of CMB data, known
as the preference
for ``negative''
effective neutrino masses \cite{Craig2024:neg_nu,Green:2024xbb,GreenMeyers2025:neg_nu_mass}. 
This tension is driven
by the CMB lensing 
4-point function reconstruction, which prefers a larger amplitude than the \textit{Planck} 2018 CMB primary prediction. 
If one formally allows  
the sum of neutrino masses to be negative (in terms of their effect on 
matter perturbations), 
they produce 
an enhancement of the
matter power spectrum instead of a suppression (due to free-streaming), which produces the desired excess of 
CMB lensing power. 
The negative neutrino 
mass preference 
increases upon addition of the DESI BAO data. In the 
$\Lambda$CDM model,
the DESI BAO measurements imply a lower value of $\Omega_m$, which includes the contribution from neutrinos, proportional to their mass. The negative neutrino masses 
in this case account
for the DESI's $\Omega_m$ 
deficit with respect to \textit{Planck} 2018.

We now consider whether the 
preference for effective negative neutrino masses holds for 
a range of CMB likelihoods, which have varying levels of evidence in favor of the lensing anomaly.
Ref.~\cite{GreenMeyers2025:neg_nu_mass} showed that the preference for negative neutrino masses
exists in both the PR4 likelihood of Ref.~\cite{Tristram2024:pr4} as well as in \textit{Planck} 2018. 
We show here that 
this is also the case for the alternative CS+ACT and W+ACT likelihoods that we describe in Section~\ref{sec:data}.
The key observation here 
is that even though the
``lensing excess'' tension
varies or 
disappears altogether 
in these
likelihoods relative to \textit{Planck} 2018, 
their high $\Lambda$CDM-inferred $\Omega_m$ values
are still in tension
with the BAO, which 
gives rise to the preference for the ``negative'' neutrino masses. 

Our analysis is similar
to that of~\cite{GreenMeyers2025:neg_nu_mass}, but is based
on CS+ACT 
and
W+ACT likelihoods. 
We combine these with 
DESI BAO in the analyses presented in Fig.~\ref{fig:AL_mnu_Omm}, Fig.~\ref{fig:negnumass_1d}, and Tab.~\ref{tab:al_mnu_alt_like}.
To allow for negative effective neutrino masses, we follow the linear extrapolation scheme of Ref.~\cite{Sailer2026:hightau}, i.e. we evaluate all observables at two points: $\sum  m_\nu = 0$ and $\sum  m_\nu = |\sum m_{\nu,\mathrm{eff}}|$. The negative-mass prediction through $C_\ell(-|m|) = 2\,C_\ell(0) - C_\ell(|m|)$ corresponds to a first-order Taylor expansion around $\sum m_\nu = 0$ which is accurate enough for the small masses explored by the current data. For the positive neutrino mass case, we assume three degenerate massive states \cite{planck18}. We sample $\sum m_{\nu,\mathrm{eff}}$ with a flat prior over $[-1, 1]\,$eV. 

Figure~\ref{fig:negnumass_1d} shows the resulting one-dimensional marginalized posteriors on $\Sigma m_{\nu,\mathrm{eff}}$. 
We see that the baseline CMB + low-$\ell$ EE posterior
is peaked in the negative
mass region for both likelihood combinations. The oscillation floor $0.06$eV is disfavored 
by both CS+ACT and 
W+ACT 
at the level of $\approx 2\sigma$.
Replacing the $\texttt{SRoll2}$ low-$\ell$ EE likelihood with the higher-$\tau$ prior ($\tau=0.11 \pm 0.006$) shifts the posteriors back towards $\Sigma m_{\nu,\mathrm{eff}} \gtrsim 0$ for both likelihoods, indicating that the apparent pull towards negative effective neutrino mass is driven primarily by the low-$\ell$ EE constraint on $\tau$ rather than by the choice of high-$\ell$ CMB likelihood. 

\begin{figure*}
    \centering
    \includegraphics[width=0.49\linewidth]{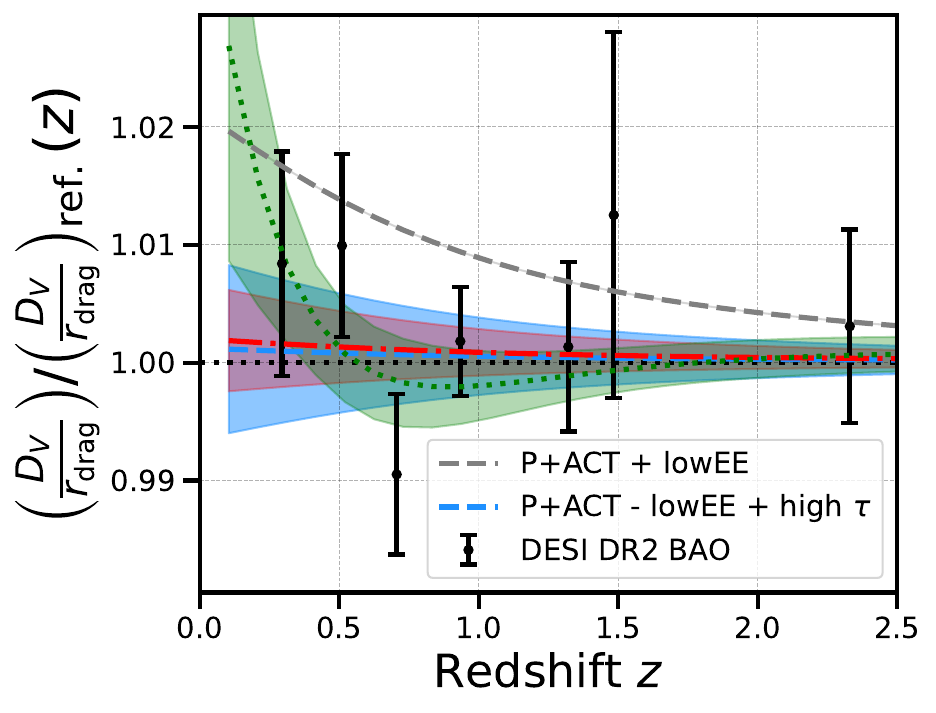}
    \includegraphics[width=0.49\linewidth]{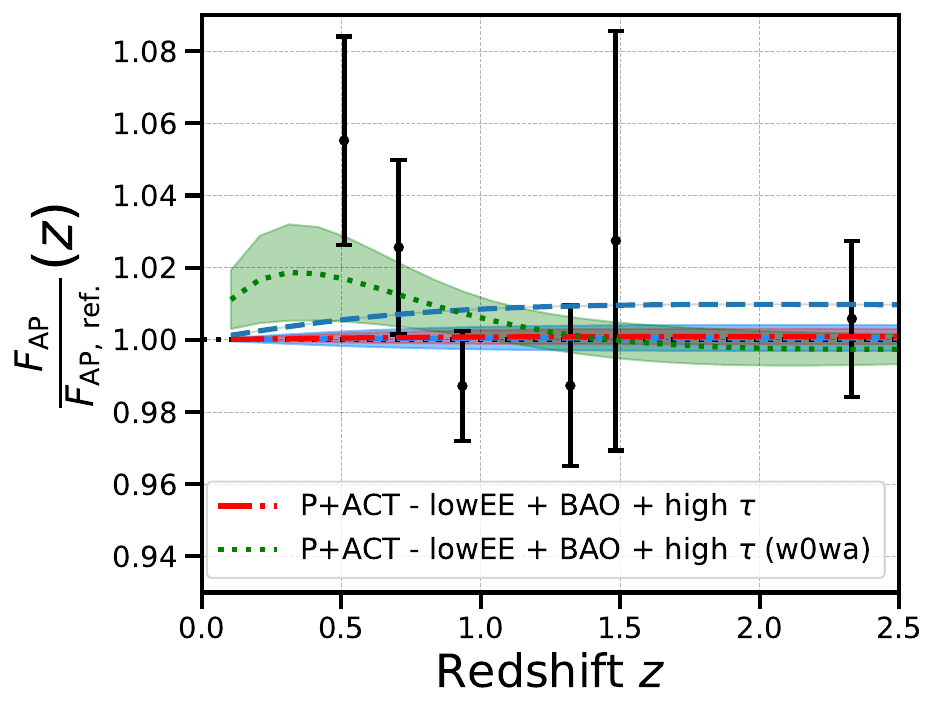}
    \caption{\textbf{Reduced BAO distance scales:}
    \textit{Left}: Isotropic BAO distance ratio posterior predictions for several models (colored curves and bands) compared to DESI data (black points). 
    \textit{Right}: Similar results for the BAO  $F_{AP}$ ratio. 
    In both panels, we divide the distance ratios by their value in a reference cosmology, here taken to be the best-fit cosmology in $\Lambda$CDM in the high-$\tau$ model using CMB data only (without low-$\ell$ EE polarization).
    Without fitting, the CMB high-$\tau$ model (blue) agrees nearly perfectly with the BAO data, a significant improvement over the + low EE baseline (gray dashed).
    When including BAO data in the analysis (red dash-dotted) the results barely change from the CMB high-$\tau$ case (blue).
    The addition of variation of evolving dark energy ($w_0w_a$, green) when including BAO data barely improves the agreement with the data.
    }
    \label{fig:dvrdbao}
\end{figure*}

To understand this result, 
we inspect the corner
plot shown in the right panel of Fig.~\ref{fig:AL_mnu_Omm}.
Here we see that combining CMB with BAO data sets $\Omega_m$, which allows for negative neutrino mass when low-$\ell$ EE data is included, but not in the high-$\tau$ Universe.
The last four rows of Tab.~\ref{tab:al_mnu_alt_like} show that the effective
neutrino mass model allows $\Omega_m\approx 0.296$ when $\sum m_{\nu,\mathrm{eff}}$ takes negative values,
which is 
the value of $\Omega_m$ preferred by the BAO.
This is lower than the value of $\Omega_m$ inferred
from these likelihoods in the absence of the BAO
($\Omega_m\approx 0.32$, 12th-16h rows of Tab.~\ref{tab:al_mnu_alt_like}).
We see that the  
combination of  CMB+BAO data 
clearly prefers 
a lower value of $\Omega_m$,
alleviating the tension between the BAO and CMB in terms of $\Omega_m$.
See Fig.~\ref{fig:nmnu_omm_nodesi} and Appendix~\ref{sec:almnu_triangle} for further discussion of this point.

Remarkably, our 
high-$\tau$ Universe 
predicts a
lensing signal with an amplitude that is 
consistent with the  
4-point function lensing contribution,
This is possible because when $\tau$ is high, $A_s$ shifts rather than $\Omega_m$ to accommodate this contribution. 
This supports a lower value of $\Omega_m$,
which blocks 
``negative'' neutrino mass.

\subsection{BAO Distance Ratio Concordance \label{subsec:bao_distance}}

The recent DESI BAO results \cite{desibao2024,AbdulKarim2025:desi_dr2_bao} indicate that the distances to low redshift galaxies are underpredicted by the $\Lambda$CDM model that best fits CMB data.
The high-$\tau$ model reduces the CMB prediction for low-redshift distances, bringing the two datasets into agreement.
We show this explicitly in Figure~\ref{fig:dvrdbao}.
There, the high-$\tau$ model posterior inferred from CMB data without the low-$\ell$ EE polarization likelihood is shown (in blue), which passes almost directly through the DESI DR2 data points. 
We stress that high-$\tau$ prediction is not a fit, yet agrees extremely well with the data.
To compare, the CMB + low-$\ell$ EE likelihood overprediction of the distances is also shown (in gray).
Compared to the baseline case of CMB+BAO with the low EE contribution removed, we find  a $\Delta\chi^2 = \chi^2_{\mathrm{P+ACT-low~EE+BAO}+\mathrm{h}.\tau.}-\chi_{\mathrm{(P+ACT+low~EE)+BAO-low~EE}}^2=-5.0$\footnote{All quoted $\Delta\chi^2$ values for CMB+BAO will be quoted with respect to this $\Lambda$CDM reference}.

The high-$\tau$ reconciliation of these two datasets, the BAO and the CMB, is possible because in the high-$\tau$ model, the CMB-inferred $\Omega_m$ decreases significantly (see Tab.~\ref{tab:main}, discussion in Sec.~\ref{subsec:cmb}) and $H_0$ also increases. 
While we did use the DESI-based $H_0$ to infer the high-$\tau$ prior mean, we did not specifically adjust $\Omega_m$.
The latter is only weakly correlated with $H_0$ in DESI-full-shape data, so the agreement we see in Fig.~\ref{fig:dvrdbao} is a genuine new test of the predictions of the high-$\tau$ Universe.

\subsubsection{Neutrino mass \label{subsec:nu_mass}}

The high-$\tau$ scenario allows for a tentative detection of positive neutrino mass from the combination of CMB and BAO data.
Following Ref.~\cite{planck18}, we approximate the neutrino mass sector with three degenerate massive states when considering the $\Lambda$CDM model with varied neutrino mass $\sum m_\nu$.
In a high-$\tau$ Universe, we find evidence for the neutrino mass sum at $2\sigma$, 
\begin{equation}
    \sum m_\nu = 0.10_{-0.05}^{+0.04}~\text{eV}.
    \label{eqn:mnu_bao_const}
\end{equation}
Figure~\ref{fig:mnu} shows that this result holds regardless of our use of ACT or SPT data (including different lensing data).
This finding is consistent with the results of Ref.~\cite{Jhaveri2025:hightau,Sailer2026:hightau,Allali2026:reion_allali_H0_1}, who find that removing \textit{Planck} low-$\ell$ EE data allows a positive posterior correlation between $\tau$ and $\sum m_\nu$.
For the case of free neutrino mass sum, we find $\Delta \chi^2 = -4.7$.
Ref.~\cite{LoverdeWeiner2024:BAO} also found significant evidence for $\sum m_\nu >0$ when combining CMB and DESY5 SNe data (in their Fig.~12); though the large value of $\Omega_m$ preferred by DESY5
is in tension with the inferred value from DESI BAO\footnote{Elsewhere in this work, we consider the updated DES Dovekie SNe sample, which prefers lower $\Omega_m$ in $\Lambda$CDM.}. 
Here we separately find evidence for $\sum m_\nu >0$ from the CMB+BAO alone.
We discuss the implications of an especially high value of $\tau$ for ``negative'' neutrino mass \cite{GreenMeyers2025:neg_nu_mass} in Section~\ref{subsec:lens_neg_nu_mass}.

\subsubsection{Dark energy and $w_0w_a$CDM}
\label{subsec:lcdm_cmb_bao_sne_w0wa}

We now turn to the combination of CMB, DESI BAO, and optionally, SNe data in the context of a dynamical dark energy model. 
The BAO and SNe data are in some tension within $\Lambda$CDM, which might lead to some significant cracks
in the high-$\tau$ $\Lambda$CDM model. 
Here we show, however, that these cracks do not emerge: the addition of the SNe data does not lead to a significant preference for $w_0w_a$ model over the cosmological constant (C.C.).

\begin{figure}
    \centering
    \includegraphics[width=\linewidth]{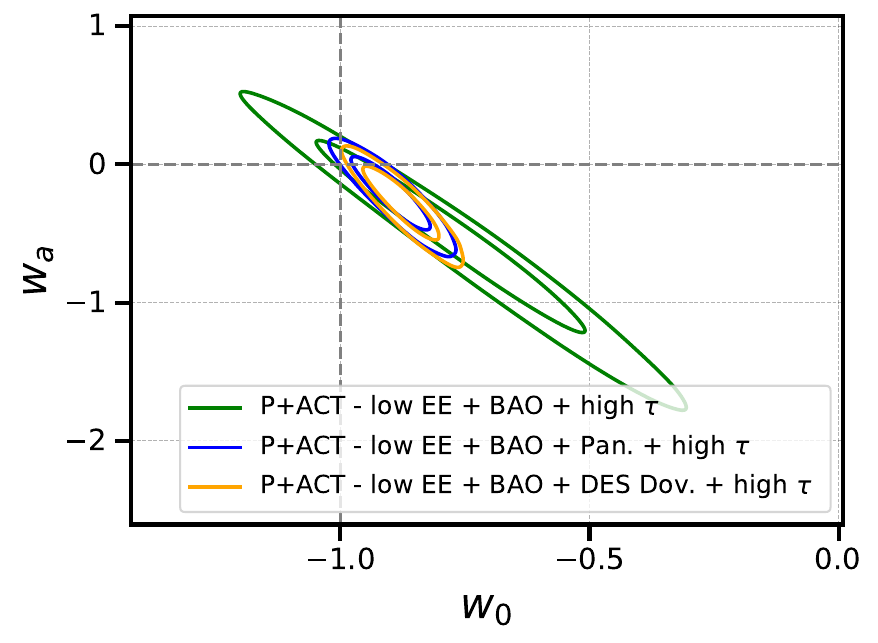}
    \caption{\textbf{No evolving dark energy:}
    We find no evidence for evolving dark energy from our fiducial CMB - BAO combination (P+ACT + DESI) in the high-$\tau$ Universe with the $w_0w_a$ model (1$\sigma$ consistency, green contour). 
    When adding uncalibrated Type Ia SNe magnitude measurements from Pantheon+ (blue contour) and DESY5 Dovekie (orange contour), we find $<2\sigma$ evidence for evolving dark energy. 
    }
    \label{fig:w0wa}
\end{figure}

When considering the combination of CMB-high-$\tau$  and DESI DR2 BAO data in an evolving dark energy model ($w_0w_a$CDM), we find no evidence for evolving dark energy
\begin{align}
    w_0 &= -0.761_{-0.192}^{+0.166} \\
    w_a &= -0.56_{-0.41}^{+0.51},
    \label{eqn:w0wa_cmb_bao_const}
\end{align}
with less than a 1$\sigma$ difference from the C.C. values of $w_0=-1,w_a=0$ in the $w_0-w_a$ plane (see Fig.~\ref{fig:w0wa}).
This lack of preference for evolving dark energy reflects the agreement between the CMB and BAO distance ratios demonstrated in Fig.~\ref{fig:dvrdbao}.
Within $w_0 w_a$CDM, we find $\Delta \chi^2=-7.8$ in the high-$\tau$ Universe.
The posterior distances in the $w_0-w_a$ model (green band) provide no significant improvement over that of the high-$\tau$ $\Lambda$CDM model (red band [CMB+BAO], nearly coincident with blue band [for CMB alone]).
This result is consistent with the interpretation that the increased preference for non-C.C. $w_0-w_a$ when low-$\ell$ EE data is included is due to $w_0$ and $w_a$ counteracting the too-high $\Omega_m$ preferred by the CMB.
We also find no preference for non-zero curvature within the high-$\tau$ Universe, making its addition to solve the CMB-BAO tension in the value of $\Omega_m$ unnecessary (see Appendix~\ref{subsec:curvature}), as increasing $\tau$ already allows us to lower the CMB-predicted BAO distance ratios to agree with the data.

We also consider combining CMB and BAO data also with uncalibrated Type Ia supernovae (SNe) magnitude measurements.
We find that when additionally combining with the Pantheon+ supernovae sample, the C.C. values are within the $2\sigma$ contour of our high-$\tau$ constraint, with marginalized constraints of
\begin{align}
    w_0 &= -0.895_{-0.052}^{+0.051} \\
    w_a &= -0.22_{-0.16}^{+0.19},
    \label{eqn:w0wa_cmb_bao_snepp_const}
\end{align}
thereby weakening the preference for evolving dark energy by $>1\sigma$ from the comparable case with low EE data included \cite{AbdulKarim2025:desi_dr2_bao}.
Due to our high-$\tau$ prior, this provides a more dramatic version of the result already seen in Refs.~\cite{Sailer2026:hightau,Jhaveri2025:hightau,act_exts_calabrese}.
We also find that using the revised DES Dovekie Type Ia SNe sample\footnote{We do not include the Union3 SNe sample, as it has recently been revised in Ref.~\cite{Hoyt2026:union3p1}, and, as far as we are aware, the revised likelihood is not publicly available as this manuscript is being finalized.} does not significantly change this conclusion, though the posterior mean values slightly shift away from that of the cosmological constant, with
\begin{align}
    w_0 &= -0.874_{-0.050}^{+0.049} \\
    w_a &= -0.29_{-0.17}^{+0.19}.
    \label{eqn:w0wa_cmb_bao_snedd_const}
\end{align}
The only mildly decreased preference for the C.C. point when SNe are included reflects the poor constraining power of SNe alone in the $w_0-w_a$ plane - the SNe alone are consistent with the C.C. values at $1\sigma$.
It is only in combination with CMB+BAO that the likelihood products with SNe slightly upweight the region of $w_0>-1$.

Ref.~\cite{Jhaveri2025:hightau} similarly explored a scenario where $\tau$ is elevated and stated that the removal of low-$\ell$ EE (and expanded $\tau$ range) cannot resolve the tension between the CMB, BAO, and SNe, and so the preference for evolving dark energy.
With our high-$\tau$ scenario, we find that such a resolution is possible.
This is almost certainly due to our use of Pantheon+ and the revised DESY5 Doevekie sample, which, with the low-$\ell$ EE data, are much less in favor of nontrivial evolving dark energy (in $w_0-w_a$) than when considering the unrevised DESY5 sample used in Ref.~\cite{Jhaveri2025:hightau}.

\subsection{The limits of raising $H_0$ with high $\tau$  \label{subsubsec:highh0tau}}

\begin{table*}[]
    \centering
  \begin{tabular}{|c|ccccc|} \hline
    \textbf{Dataset  }  
    & $\omega_{\rm m}$
    & $H_0$ 
    & $\Omega_m$ 
    & $\sigma_8$
    & $S_8$
    \\
    \hline
    P+ACT - low-$\ell$  + v. h$\tau$ 
    & $0.1374 \pm 0.0009$ 
    & $69.88_{-0.45}^{+0.44}$ 
    & $0.281 \pm 0.005$ 
    & $0.855_{-0.04}^{0.005}$ 
    & $0.828 \pm 0.010$ 
    \\
    P+ACT - low-$\ell$ EE + v. h$\tau$ + BAO 
    & $0.1388 \pm 0.0006$ 
    & $69.21 \pm 0.28$ 
    & $0.290\pm 0.003$ 
    & $0.858 \pm 0.004$ 
    & $0.843 \pm 0.007$ 
    \\
  \hline
    \end{tabular}
    \caption{\textbf{Too-high $\tau$ drives $\Omega_m$ too low:}
    $\Lambda$CDM parameter constraints for the ``very-high-$\tau$'' scenario ($\tau=0.15\pm0.006$) with CMB (P+ACT) data alone and from the combination of CMB+BAO data.
    The very-high-$\tau$ result pushes $H_0$ higher yet moves $\Omega_m$ lower (cf. Tab.~\ref{tab:main}), traveling along the CMB degeneracy.
    We see that the inclusion of BAO data in this model lowers $H_0$ due to the breaking of the CMB degeneracy, as discussed in Fig.~\ref{fig:noH0soln}.
    The value of $S_8$ also increases significantly with the addition of BAO data, which is driven by the change in $\Omega_m$.
    }
\label{tab:very_high_tau}
\end{table*}

As we motivated in the estimate of Section~\ref{sec:intro}, raising $\tau$ naively seems to be a possible way to reconcile the higher $H_0$ values inferred from local measurements as well as, albeit with a lower value, from DESI data, since it allows us to also raise $H_0$.
However, we show that while this is true to an interesting extent, this argument cannot be continued to a full resolution of the Hubble tension.

\begin{figure}
    \centering
    \includegraphics[width=\linewidth]{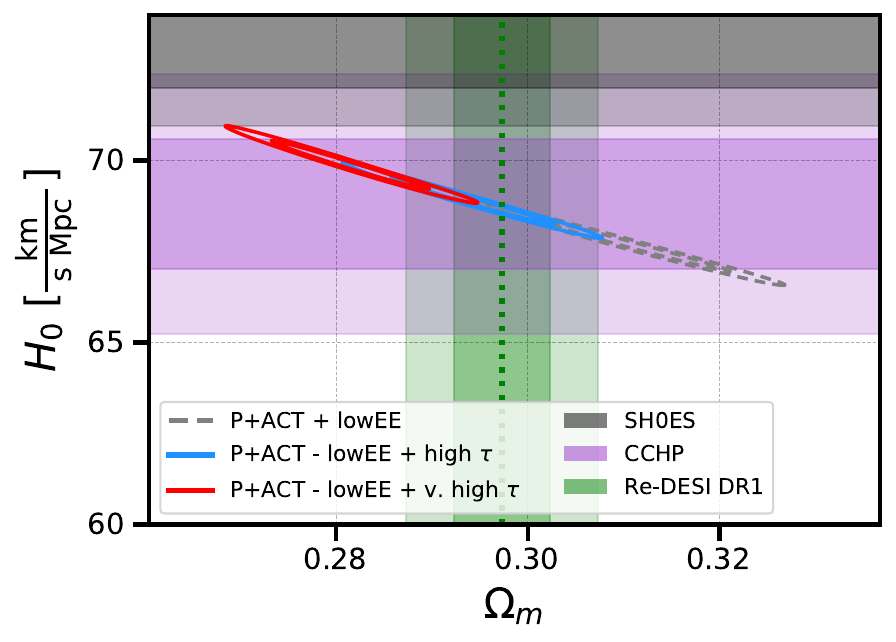}
    \caption{\textbf{Extremely high $\boldsymbol{\tau}$ cannot solve the Hubble tension:} 
    In grey (dashed) we show the baseline constraint in the $\Omega_m$-$H_0$ plane.
    In blue, we show our fiducial high-$\tau$ constraint, while in red, we show the  case of extremely high $\tau$ ($\tau = 0.15 \pm 0.006$).
    Horizontal shaded bars show $1\sigma$ and $2\sigma$ regions for local $H_0$ measurements from Ref.~\cite{Freedman2025:CCHP} and Ref.~\cite{Riess:2021jrx}.
    Vertical shaded bars (green) show $1\sigma$ and $2\sigma$ regions from the main constraints in the DESI DR1 reanalysis of Ref.~\cite{Ivanov2026:desi_dr1_4_kitchen_sink}.
    The extremely high-$\tau$ case (red) does allows for large $H_0$, significantly reducing the Hubble tension, but favors an $\Omega_m$ that is $\gtrsim3\sigma$ away from LSS result (green).
    }
    \label{fig:noH0soln}
\end{figure}

Fig.~\ref{fig:noH0soln} shows the $H_0-\Omega_m$ plane, with baseline constraints from the CMB with low-$\ell$ EE data (in gray) as well as with the high-$\tau$ prior (in blue).
We also consider an even more extreme very (v.) high-$\tau$ prior of $\tau = 0.150 \pm 0.006$ (in red, see also Tab.~\ref{tab:very_high_tau}).
While the CMB alone allows the v. high $\tau$ model to push up to $H_0 \approx 70$ km/s/Mpc along the CMB $\Omega_m h^3$ degeneracy, tantalizingly moving towards a meaningful resolution of the tension with SH0ES (gray band), this result is in $>2\sigma$ disagreement with the DESI constraints on $\Omega_m$ (green band, result from Ref.~\cite{Ivanov2026:desi_dr1_4_kitchen_sink}).

This implies that $H_0$ can increase with higher $\tau$, but not enough to significantly resolve the late-Universe Hubble tension with SH0ES while simultaneously satisfying concordance with $\Omega_m$ inferred from large-scale structure.
Even discarding LSS data, since uncalibrated Type Ia SNe prefer somewhat larger values of $\Omega_m$ than the BAO, these data will further disfavor making $\tau$ extremely large as a way to further resolve the Hubble tension beyond what we achieve in our baseline high-$\tau$ case.
Accordingly, further slackening of the Hubble tension with SH0ES\footnote{We note that our results are perfectly consistent with the wider Carnegie-Chicago Hubble Program (CCHP) errorbars (purple band in Fig.~\ref{fig:noH0soln}) \cite{Freedman2025:CCHP}} may remain out of reach without new forms of resolution.

\begin{figure*}
    \centering
    \includegraphics[width=0.95\linewidth]{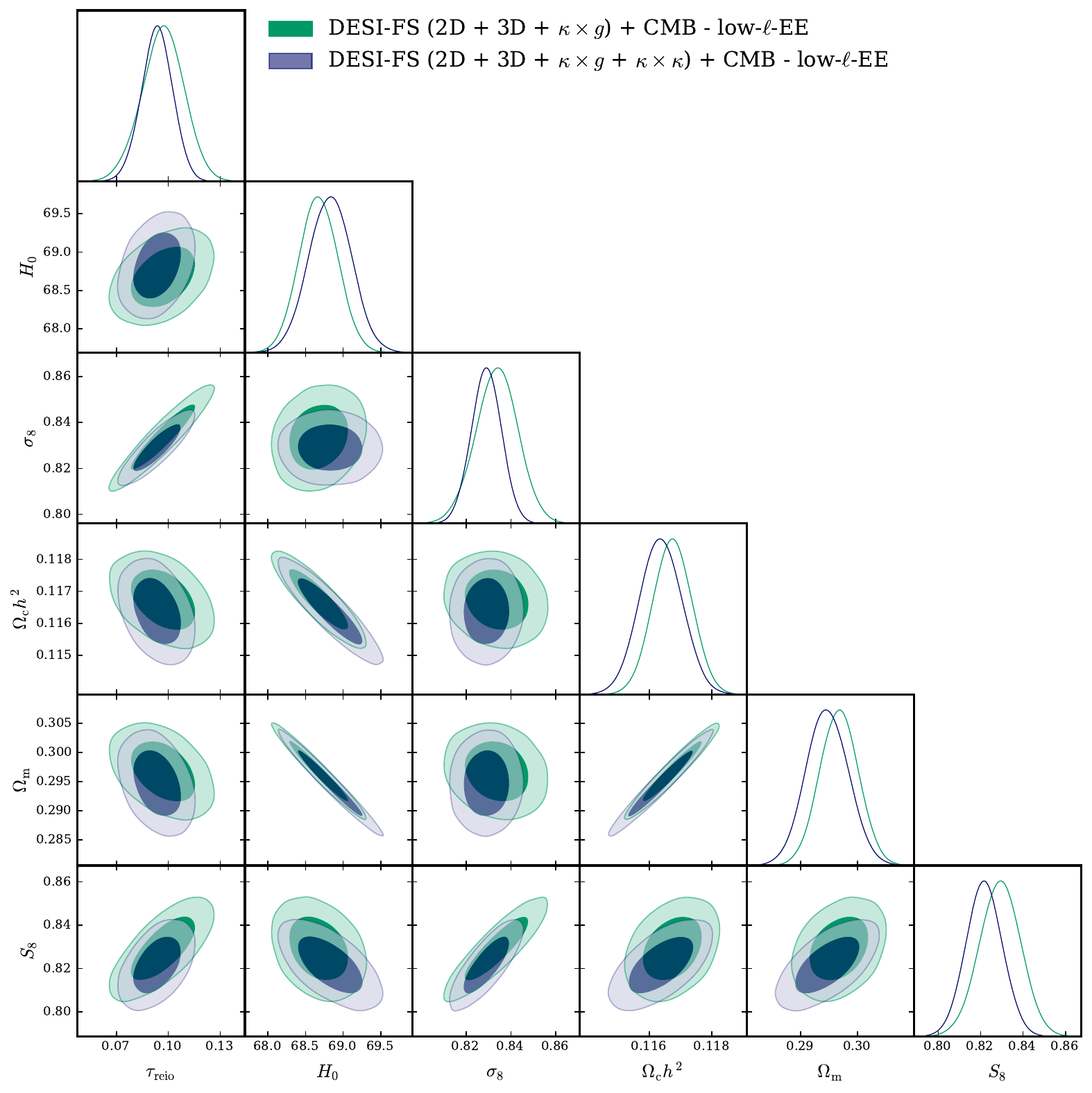}
    \caption{\textbf{DESI full shape favors a high-$\tau$ Universe:} Joint posteriors on $(\tau,\,H_0,\,\sigma_8,\,\omega_\mathrm{cdm},\,\Omega_m,\,S_8)$ from the free-$\tau$ $\Lambda$CDM analysis combining the DESI full-shape compressed Gaussian likelihood with CS(PR4)+ACT - \textit{Planck} PR4 NPIPE CamSpec high-$\ell$ TTTEEE, ACT DR6 CMB-only, and the \textit{Planck} 2018 commander low-$\ell$ TT (no low-$\ell$ EE; $\tau$ is sampled with a wide uniform prior); the triangle shows $\omega_\mathrm{cdm}$, while the derived $\omega_m \equiv \Omega_m h^2$ is reported in Table~\ref{tab:DESI_FS_tau}. Green contours show DESI FS (2D + 3D) which include the lensing-galaxy cross-correlation and blue contours additionally include the lensing auto-correlation (and their covariances).
    }
    \label{fig:lcdm_triangle_DESIFS}
\end{figure*}

It is interesting to compare the performance of the high-$\tau$ model with beyond-$\Lambda$CDM proposals to address the Hubble tension. 
In these models, a larger value of $H_0$ inferred from the CMB can be achieved by new physics that modifies the expansion history (see, e.g., Refs.~\cite{DiValentino2021:H0review,h0_olympics,Abdalla:2022yfr} for reviews of solutions to the $H_0$ tension 
).
One point of comparison is the early dark energy model (EDE), see Ref.~\cite{McDonough:2023qcu} for a review of observational constraints.
Ref.~\cite{act_exts_calabrese} finds $69.96^{+0.8}_{-1.5}$ km/s/Mpc 
in the $n=3$ EDE model  using their ``P-ACT-LB'' combination. 
This corresponds to a combination we consider here - \textit{Planck}+ACT \texttt{plik} with appropriate cuts and combined lensing along plus DESI DR2 BAO.
In this case, we find in the high-$\tau$ model that $H_0 = 68.83_{-0.26}^{+0.27}$ km/s/Mpc within $\Lambda$CDM. 
A second point of comparison is a model with varying electron mass, yielding $H_0=70.4\pm 1.1$
km/s/Mpc
\cite{Baryakhtar2024:vary_me}.

In both of these models the optimal $H_0$ values are not very different from our $\Lambda$CDM constraint. 
An alleviation of the tension in these two models is achieved primarily by means of enlarging the error-bars. 
Nevertheless, the varying electron mass cosmology is in $2\sigma$ tension with respect to the most recent local $H_0$ determination. Our high-$\tau$ model is in $4.4\sigma$ tension with this result, so some new physics might still be needed in order to account for the Hubble tension.
Our high-$\tau$
proposal can make this 
job easier as it 
suggests
a higher $H_0$
value as a $\Lambda$CDM starting point for models 
that address the tension. 
Refs.~\cite{Allali2026:reion_allali_H0_1,Allali2025:tau_H0_lowEE}
showed, however,
that this approach 
does not work 
for the EDE and the dark radiation models,
where freeing $\tau$ by dropping low$\ell$ EE \textit{Planck} data leads only to a mild 
additional 
reduction of the 
$H_0$ tension.
Nevertheless, it would be interesting to consider this
possibility 
for other models, e.g.
the varying election mass, to see if the
combination of new physics and 
high $\tau$ values 
can help alleviate 
the Hubble tension
more successfully 
than each 
proposal separately.

\subsection{$\tau$ from DESI full-shape data \label{subsec:fullshape}}

In Section~\ref{sec:intro}, 
we motivated our choice of 
the concordance 
$\tau$ value by the
CMB-independent 
cosmological constraints from 
DESI EFT-based 
full-shape  analyses.
In this Section
we extract a consistency-based value of $\tau$ from the direct combination of CMB and DESI-FS data. In this analysis we fix the 
neutrino mass to the oscillation floor $\sum m_\nu  = 0.06$
eV.

In principle, the full-shape data 
provides additional channels
to probe $\tau$
through its correlation with $\omega_m$ and $H_0$ extracted from the shape of the matter transfer function,
and the equality
scale, respectively.
Therefore, one might expect this data 
to affect the 
inference of 
$\tau$ from 
the CMB and BAO. 

\begin{table*}[t]
\begin{ruledtabular}
\begin{tabular}{lccccc}
Dataset
 & $\tau_\mathrm{reio}$
 & $H_0$
 & $\Omega_\mathrm{m}$
 & $\sigma_8$
 & $S_8$ \\
\hline
DESI-FS (2D$+$3D$+\kappa\times g$)
 & $0.097\pm 0.012$ & $68.67\pm 0.26$ 
 & $0.2968\pm 0.0034$ & $0.8337\pm 0.0095$ & $0.8291\pm 0.0098$ \\
DESI-FS (2D$+$3D$+\kappa\times g+\kappa\times\kappa$)
 & $0.0934\pm 0.0090$ & $68.83\pm 0.28$ 
 & $0.2947\pm 0.0037$ & $0.8290\pm 0.0067$ & $0.8217\pm 0.0086$ \\
\end{tabular}
\end{ruledtabular}
    \caption{\textbf{DESI full-shape prefers $\tau\geq 0.09$:}
    $\Lambda$CDM parameter constraints with fixed neutrino masses ($\Sigma m_{\nu}=0.06\, \mathrm{eV}$) and varying the optical depth parameter using CMB (P+ACT) data paired with DESI full-shape information. All fits exclude the low-$\ell$ EE data. 
    In the first row the CMB-lensing information enters through the FS likelihood itself via the lensing-galaxy cross-correlation $\kappa\times g$ and, in the second row, the
    lensing auto-spectrum $\kappa\times\kappa$ where both include an analytic cross-covariance~\cite{deBelsunce:2025qku}. The Hubble constant is given in units of $[\mathrm{km\,s^{-1}\,Mpc^{-1}}]$. The values for the optical depth, when including the $\kappa\times g$ cross-correlation and covariance,  yield $\tau = 0.097\pm 0.012$ and, when additionally including the lensing auto-correlation, $\tau = 0.0934\pm 0.0090$ which are $-1.1$ and $-1.8\sigma$ lower than our high-$\tau$ prior. The corresponding 1D and 2D marginalized posterior contours are shown in Fig.~\ref{fig:lcdm_triangle_DESIFS}.
    }
    \label{tab:DESI_FS_tau}
\end{table*}

The DESI FS ``kitchen sink'' (2D + 3D) chains from Ref.~\cite{Ivanov2026:desi_dr1_4_kitchen_sink} are sampled using priors on $\omega_b\!\sim\!\mathcal{N}(0.02237,0.00015^2)$ and $n_s\!\sim\!\mathcal{N}(0.9649,0.0042^2)$ which we remove in post-processing by re-weighting the chains. Note that the DESI FS 2D + 3D chains are based on the 
one-loop power spectrum 
and one-loop bispectrum
of DESI galaxies from~\cite{Chudaykin:2025aux}.
We combine both compressed FS likelihoods with CMB primary data from PR4 CamSpec high-$\ell$ TTTEEE, ACT DR6 CMB-only,  \textit{Planck} 2018 \textsc{Commander} low-$\ell$ TT (the CS(PR4)+ACT combination, see Section~\ref{sec:data}).
We account for the cross-covariance between the 2D and 3D full-shape samples; first only with the PR4 and DR6 lensing cross-spectrum and, second, with the lensing auto-spectrum by computing an analytic Gaussian covariance~\cite{GarciaGarcia19, deBelsunce:2025qku} and the joint constraints are listed in Table~\ref{tab:DESI_FS_tau} and shown in Fig.~\ref{fig:lcdm_triangle_DESIFS}.
We have verified that the Gaussian posteriors match the full 1D marginalized posteriors. 

Leaving $\tau$ free with a wide uniform prior $\tau\sim\mathcal{U}(0.01,0.2)$ and fixing $\sum m_\nu=0.06$~eV, the joint fit yields, 
for the DESI FS (2D + 3D) configuration including the lensing-galaxy cross-correlation, $\tau=0.097\pm0.012$; 
and for the DESI FS (2D + 3D) configuration including the lensing auto-correlation as well as lensing-galaxy cross-correlation, $\tau=0.0934\pm0.0090$.
We note that the inclusion of the $\kappa\times g$ spectra pulls the $\tau$ posterior to $-1.1\sigma$ agreement with our high-$\tau$ prior. 
Further adding the lensing auto-correlation (and the corresponding covariance matrix) leads to a substantially tighter constraint on $\tau$ that sits between the previous best-fit $\tau$ values of the chains and is $-1.8\sigma$ lower than the central value of our prior.  

The FS value of $\tau$
is somewhat higher, but fully consistent with the 
CMB+BAO result of~\cite{Sailer2026:hightau}
$\tau=0.090\pm 0.012$. We see that the DESI DR1 full-shape data does not
substantially affect the inference of $\tau$  from LSS and the CMB, but this may change in the future with more precise full-shape data from DESI and \textit{Euclid}.

While the value we recover in this case is somewhat lower than the central value of the prior that we consider, we note that in all FS analyses we kept $\sum m_\nu$ fixed. A more general analysis should include varying both 
$\tau$ and 
the neutrino mass
in the CMB+FS analysis.
However, in this case the 
compression of the FS likelihood to $(H_0,\omega_{\rm cdm},\sigma_8)$
is not expected to be accurate, 
and a full analysis is required.
In addition, it would be interesting to infer $\tau$ from other LSS datasets, 
e.g. DES $3\times 2$pt correlations~\cite{DES:2026fyc}.
We leave these investigations for future work.

\section{Discussion \label{sec:disc}}

The low optical depth to reionization $\tau$ preferred by \textit{Planck} 2018 data suggests unphysical cosmological neutrino mass constraints.
This suggestion is even more alarming when combining CMB data with DESI DR2 Baryonic Acoustic Oscillation distance measurements, which increases this unphysical preference.
It is known that the value of $\tau$ and neutrino mass are connected in this context, in particular through the \textit{Planck} low-$\ell$ EE polarization data \cite{LoverdeWeiner2024:BAO,Green:2024xbb}.
Recently,  
Refs.~\cite{Sailer2026:hightau,Jhaveri2025:hightau} showed that 
without the inclusion of this data, 
these most recent datasets support a
higher value of 
$\tau$ with broader uncertainty than is favored by the low-$\ell$ CMB EE polarization likelihood: $\tau = 0.09 \pm 0.012$ ($\tau=0.091 \pm 0.011$). 

Here we have simply taken this logic and pushed it further, considering what the consequences of an even higher value of $\tau$ would be if it were known \textit{at a \textit{Planck} 2018 level of confidence}\footnote{In the context of the future measurement of $\tau$, e.g., from \textit{LiteBIRD}, CLASS, or PICO, this is actually a conservative statement, as the uncertainty on $\tau$ from such an experiment will be tighter.}.
From this supposition, it follows that not only are these tensions
simply reduced, but, with the exception of some high local $H_0$ measurements, they are removed altogether at the $<2\sigma$ level.

We extend this method of relaxing the tension between these datasets, and, in summary:
\begin{enumerate}
    \item we infer the first $2\sigma$ evidence for a positive neutrino mass within $\Lambda$CDM,
    \item we remove evidence for dynamical dark energy obtained from DESI BAO measurements,
    \item we show that the \textit{Planck} CMB data (without low-$\ell$ EE polarization) shows no preference for excess lensing amplitude,
    \item we eliminate the preference for negative effective neutrino mass,
    \item we resolve the \textit{Planck} 2018 lensing anomaly through the inference of physical neutrino mass,
    \item we lower the CMB-inferred value of $\Omega_m$ to better agree with the value inferred from the BAO,
    \item we reduce the Hubble tension as seen by SH0ES to $\approx 4\sigma$ (with respect to CCHP, there is no tension) and alleviate the disagreement between DESI full-shape + CMB lensing cross-correlation data,
    \item we perform a consistency analysis using DESI full-shape 
    and BAO 
    data for the first time, which supports the high $\tau$ Universe.
\end{enumerate}

The high value of $\tau$ that we have chosen is significantly larger than not just the low-$\ell$ EE value, but also than values found in the literature of astrophysical inference of $\tau$, such as from Lyman-$\alpha$ damping wings and high-redshift James Webb Space Telescope (JWST) galaxies \citep[e.g.,][]{Umeda2024:JWST_reion,Kageura:2026ryq,Mason2026:JWST_damping}, or the Lyman-$\alpha$ forest \cite{Garcia-Gallego2025:lyaf_tau}. 
However, while these measurements are consistent with low $\tau$ under standard, rapid, single-transition reionization models, the total lack of high-redshift $z>12$ observations still allows for the possibility of a more complicated reionization history that produces significantly larger $\tau$ (e.g., a modified version of the double-reionization model of Ref.~\cite{Cen2003:double_reion}; for a recent model proposal in this direction, see Ref.\cite{Aggarwal2026:popiii_baocmb}).
Though such scenarios are not widely modeled by state-of-the-art numerical simulations of reionization \cite{GnedinMadau2022,Kannan2022:thesan,Puchwein2023:sherwood_reion,Lumina2026:Zier}, they are not explicitly ruled out by current astrophysical data. 
Complementary information as to the value of $\tau$ may also come from the future kinetic Sunyaev-Zel'dovich reionization signal \cite{2021Alvarez:kSZtau,Cain2025:ksz_reion,Beringue2025:act_dr6_foreground} or 21cm experiments \cite{Liu2016:21cm_tau}, though not at the level of precision of \textit{Planck}, although they are not expected to reach the precision of \textit{Planck}, the planned space-based \textit{LiteBIRD} mission ($\sigma(\tau)\approx 0.002$ \cite{litebird:2023}), the ground-based Cosmology Large Angular Scale Surveyor (CLASS, $\sigma(\tau)\approx 0.003$ \cite{2014SPIE.9153E..1IE}), or the previously proposed PICO ($\sigma(\tau)\approx 0.002$ \cite{pico2019}).

Regardless of the state of the astrophysical $\tau$ literature, future SNe from upcoming transient surveys (e.g., those associated with Rubin Observatory \cite{LSSTDESC2018}, and possibly at higher redshift with the Roman space telescope \cite{Roman2015})  will provide additional confidence as to whether the current preference for higher $\Omega_m$ in $\Lambda$CDM from uncalibrated Type Ia SNe magnitudes is resolute.
A persistently large value of $\Omega_m$ with shrinking uncertainties would reduce confidence in a high-$\tau$ model as a compelling solution to the CMB-BAO disagreement on $\Omega_m$ in $\Lambda$CDM. 
The reverse would lend credence to the high-$\tau$ solution, as all probes (CMB, BAO, SNe) would then agree on $\Omega_m$ at the $\sim 1 \sigma$ level.
These conclusions hold for CMB lensing data from both \textit{Planck}+ACT DR6 as well as for SPT-3G.
Tighter large-scale structure consistency constraints on $\tau$ may also be accessible through the use of galaxy-lensing cross correlations (e.g., 3x2pt) or through the full-shape 
galaxy, quasar, and Lyman-$\alpha$ clustering data. These probes 
will have the potential 
to determine the fate of the 
high-$\tau$ scenario in the
short term.

\acknowledgments
We thank 
Dan Green,
George Efstathiou, 
Simone Ferraro,
Colin Hill, 
Marc Kamionkowski, 
Marilena Loverde,
Noah Sailer,
Zachary Weiner,
and Martin White
for discussions and feedback on a draft version of this work,
as well as 
Kendrick Smith and
Matias Zaldarriaga for productive conversations, and Erik Rosenberg for help with CamSpec. 
JMS acknowledges that support for this work was provided by The Brinson Foundation through a Brinson Prize Fellowship grant.
This work is supported by the National Science Foundation under Cooperative Agreement PHY-2019786 (The NSF AI Institute for Artificial Intelligence and Fundamental Interactions, \url{http://iaifi.org/}). This research used resources of the National Energy Research Scientific Computing Center (NERSC), a U.S. Department of Energy Office of Science User Facility operated under Contract No.~DE–AC02–05CH11231. 

\appendix

\section{SPT-3G \label{subsec:spt}}
As a test of the dependence of our fiducial results on the use of \textit{Planck}+ACT primary data, we also consider a likelihood combination using SPT-3G D1 CMB data alone \cite{Camphuis2025:SPT3GD1,GeMillea2025:spt_3g_polzn}.
This data combination is somewhat less constraining than our fiducial combination, but supplies constraints that are fully consistent with our main results.
\begin{table*}[!htb]
    \centering
    \renewcommand{\arraystretch}{1.2}
    \setlength{\tabcolsep}{5pt}
    \begin{tabular}{@{}l ccccc@{}}
  \hline\hline
    \textbf{Dataset}  
    & $\omega_{\rm m}$
    & $H_0$ 
    & $\Omega_m$ 
    & $\sigma_8$
    & $S_8$
    \\
    \hline
    \multicolumn{6}{@{}l}{\textbf{$\Lambda$CDM}} \\
    \quad SPT - low-$\ell$  + h$\tau$
    & $0.1381 \pm 0.0015$ 
    & $68.99_{-0.63}^{+0.58}$ 
    & $0.290 \pm 0.008$ 
    & $0.841 \pm 0.006$ 
    & $0.828 \pm 0.015$ 
    \\
    \quad SPT - low-$\ell$ EE + h$\tau$ + BAO 
    & $0.1390 \pm 0.0008$ 
    & $68.62 \pm 0.30$ 
    & $0.295\pm 0.004$ 
    & $0.843 \pm 0.006$ 
    & $0.837_{-0.008}^{+0.009}$ 
    \\
    \multicolumn{6}{@{}l}{\textbf{$\Lambda$CDM $+ \sum m_\nu$}} \\
    \quad SPT - low-$\ell$ EE + h$\tau$ + BAO 
    & $0.1390 \pm 0.0008$ 
    & $68.42 \pm 0.38$ 
    & $0.297 \pm 0.004$ 
    & $0.834_{-0.012}^{+0.013}$ 
    & $0.830 \pm 0.012$ 
    \\
    \multicolumn{6}{@{}l}{\textbf{$w_0 w_a$CDM}} \\
    \quad SPT - low-$\ell$ EE + h$\tau$ + BAO
    & $0.1391 \pm 0.0012$ 
    & $65.94\pm 1.86$ 
    & $0.321_{-0.020}^{+0.018}$ 
    & $0.820 \pm 0.016$ 
    & $0.847 \pm 0.015$ 
    \\
    \quad SPT - low-$\ell$ EE + h$\tau$ + BAO + SNe 
    & $0.1388 \pm 0.0011$ 
    & $67.24_{-0.61}^{+0.60}$ 
    & $0.307 \pm 0.006$ 
    & $0.829 \pm 0.010$ 
    & $0.839 \pm 0.011$ 
    \\
  \hline
    \end{tabular}
    \caption{
    \textbf{SPT-3G + high $\tau$:}
    Similar to Table~\ref{tab:main}, we show $\Lambda$CDM constraints in the high-$\tau$ Universe, but here for SPT-3G CMB data.
    We see that while the constraints are more permissive, results are broadly similar.
    }
\label{tab:spt}
\end{table*}

We summarize our $\Lambda$CDM constraints with these data in Table~\ref{tab:spt}.
In brief, we find parameter constraints that are consistent with the fiducial high-$\tau$ results.
The only significant difference with SPT is that $n_s$ and $\omega_b$ are noticeably less well-constrained.
We find that, in the high-$\tau$ scenario, SPT has no preference for nontrivial $A_\mathrm{L}$ (albeit with larger uncertainties)  with
\begin{equation}
    A_\mathrm{L} = 1.00_{-0.07}^{+0.08}.
\end{equation}

When using SPT-3G data, we find extremely similar results for constraints on the neutrino mass parameter for CMB + BAO (see Fig.~\ref{fig:mnu})
\begin{equation}
    \sum m_\nu = (0.10 \pm 0.05)~\text{eV}\,,
    \label{eqn:spt_mnu_const}
\end{equation}
and evolving dark energy parameters $w_0$,$w_a$, with CMB + BAO 
\begin{align}
    w_0 &= -0.762_{-0.199}^{+0.166} \\
    w_a &= -0.56_{-0.42}^{+0.55},
    \label{eqn:spt_w0wa_const}
\end{align}
with CMB + BAO + Pantheon+
\begin{align}
    w_0 &= -0.896 \pm 0.051 \\
    w_a &= -0.21_{-0.17}^{+0.19},
    \label{eqn:spt_w0wa_const_pp}
\end{align}
and with CMB + BAO + DES Dovekie
\begin{align}
    w_0 &= -0.974 \pm 0.053 \\
    w_a &= -0.29_{-0.19}^{+0.21}.
    \label{eqn:spt_w0wa_const_dd}
\end{align}
For comparison see eqn.~\ref{eqn:mnu_bao_const} and eqns.~\ref{eqn:w0wa_cmb_bao_const}-\ref{eqn:w0wa_cmb_bao_snedd_const}, where all values are comfortably consistent at within 1$\sigma$.
This indicates that our main results are not dependent on the use of ACT or \textit{Planck} data. 

\section{Spatial curvature \label{subsec:curvature}}

When considering the combination CMB - low-$\ell$ EE + BAO (i.e., high-$\tau$ Universe) in a $\Lambda$CDM model with free spatial curvature, we infer the curvature parameters to be consistent with zero
\begin{align}
    \omega_k &= 0.00018 \pm 0.00053\\
    \Omega_k &= 0.00037_{-0.00111}^{+0.00112},
    \label{eqn:omk_omkh2_cmb_bao_const}
\end{align}
i.e., a flat Universe.
The high-$\tau$ Universe therefore has no preference for spatial curvature, which Ref.~\cite{ChenZaldarriaga2025:omk} showed can bring the CMB + low-$\ell$ EE and DESI DR2 BAO datasets into better agreement with each other in terms of $\Lambda$CDM parameters.
This is consistent with Fig.~\ref{fig:dvrdbao}, which suggests that there is no extra model freedom required beyond flat $\Lambda$CDM.

\section{Full constraints from alternative CMB likelihoods - $A_\mathrm{L}$ \& $\sum m_{\nu,\mathrm{eff}}$ \label{sec:almnu_triangle}}

\begin{figure}
    \centering
    \includegraphics[width=0.98\linewidth]{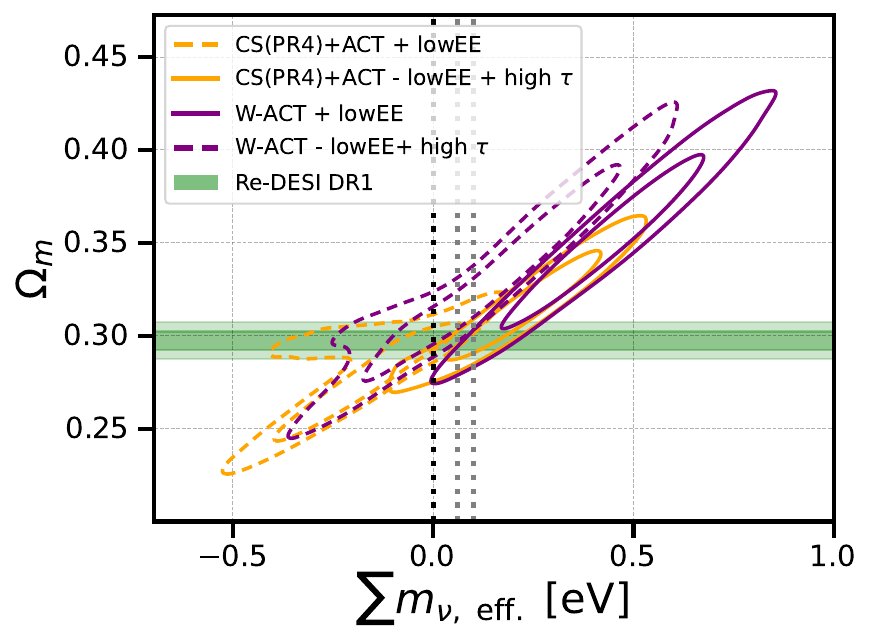}
    \caption{\textbf{DESI \& negative neutrino mass:}
    The ``negative'' effective neutrino mass model constraints, similar to the right panel of Fig.~\ref{fig:AL_mnu_Omm}, but without including DESI DR2 data.
    There is a strong degeneracy between $\Omega_m$ and the effective neutrino mass $\sum m_{\nu,\mathrm{eff}}$.
    We see that the low $\Omega_m$ favored by DESI (green band, same meaning as in Fig.~\ref{fig:noH0soln}) intersects the contours of the low-$\ell$ EE baseline models (dashed) at negative or zero effective neutrino mass values.
    In contrast, the high-$\tau$ models (solid) shift the $\Omega_m-\sum m_{\nu,\mathrm{eff}}$ degeneracy toward higher effective neutrino masses, which allows the contours to intersect\footnote{Of course, this is a low-dimensional projection of the full posterior density; however, our results in Fig.~\ref{fig:AL_mnu_Omm} are consistent with this interpretation.} in a region of positive effective neutrino mass.
    Interestingly, W+ACT has 2$\sigma$ neutrino mass preference in the high-$\tau$ model, albeit at a higher value, even without combining with DESI, while CamSpec finds slightly less significant preference for this value.
    }
    \label{fig:nmnu_omm_nodesi}
\end{figure}

Figure~\ref{fig:nmnu_omm_nodesi} shows similar contours to those presented in the right panel of Fig.~\ref{fig:AL_mnu_Omm}, but for the CMB data without the combination of DESI DR2 BAO data.
In this case, we see the strong degeneracy between $\Omega_m$ and $\Sigma m_{\nu,\mathrm{eff}}$ in both the case where low-$\ell$ EE data is included (favoring low $\tau$) and in the high-$\tau$ model.
The high-$\tau$ model shifts the contour to the right (toward positive effective neutrino masses) and slightly up for both the CS(PR4)+ACT and W+ACT likelihoods.
This result provides some indication for why our fiducial CMB likelihood as well as W+ACT and CS(PR4)+ACT show evidence for negative effective neutrino mass despite their disagreeing significance (or lack thereof) of preferences for the lensing rescaling parameter $A_\mathrm{L}$.
W+ACT + low EE in particular only prefers negative neutrino masses when combined with DESI BAO, illustrating that the variation of $\Sigma m_{\nu,\mathrm{eff}}$ chiefly allows $\Omega_m$ to decrease from the value preferred by the CMB in $\Lambda$CDM when $\tau$ is low. 

\begin{figure*}
    \centering
    \includegraphics[width=0.49\linewidth]{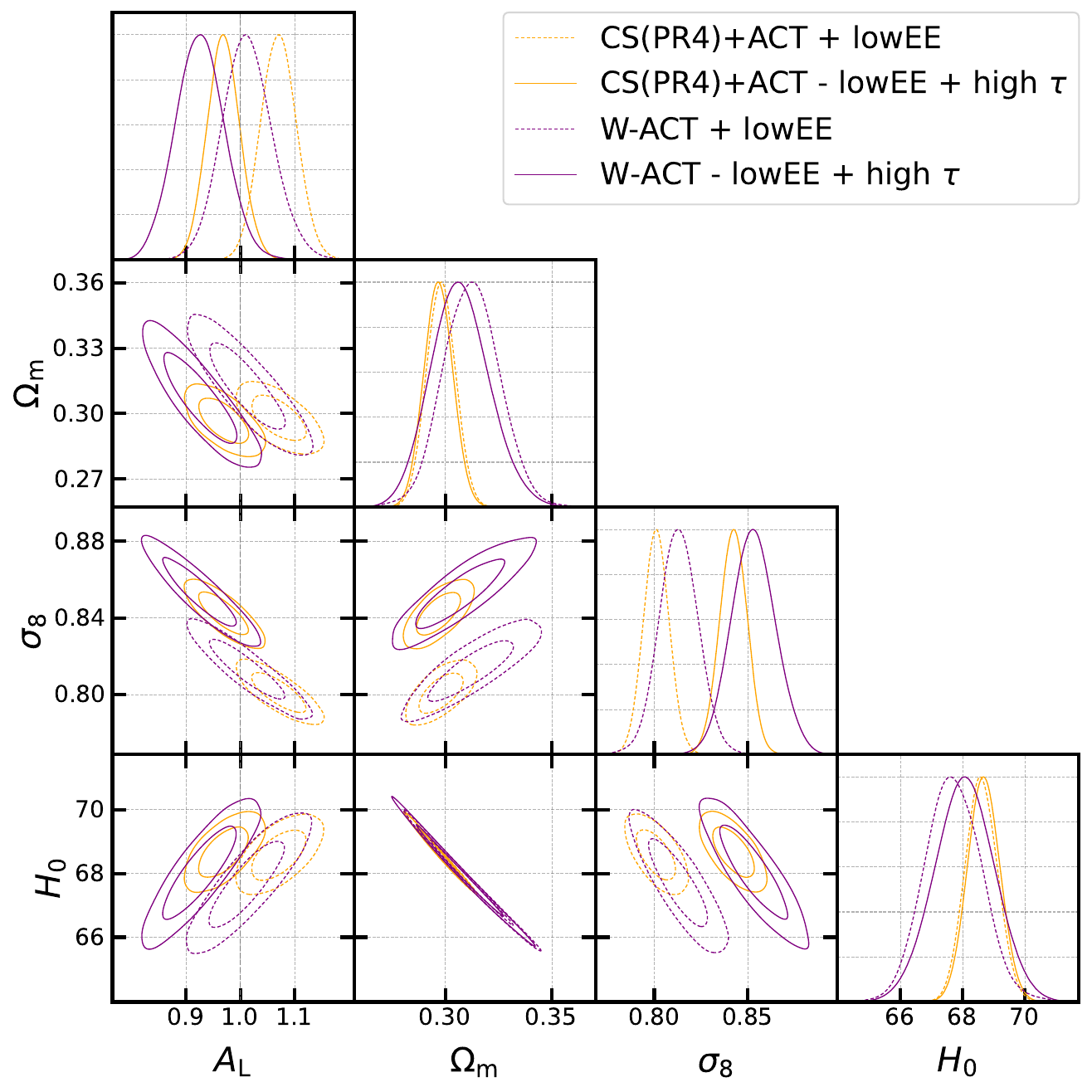}
    \includegraphics[width=0.49\linewidth]{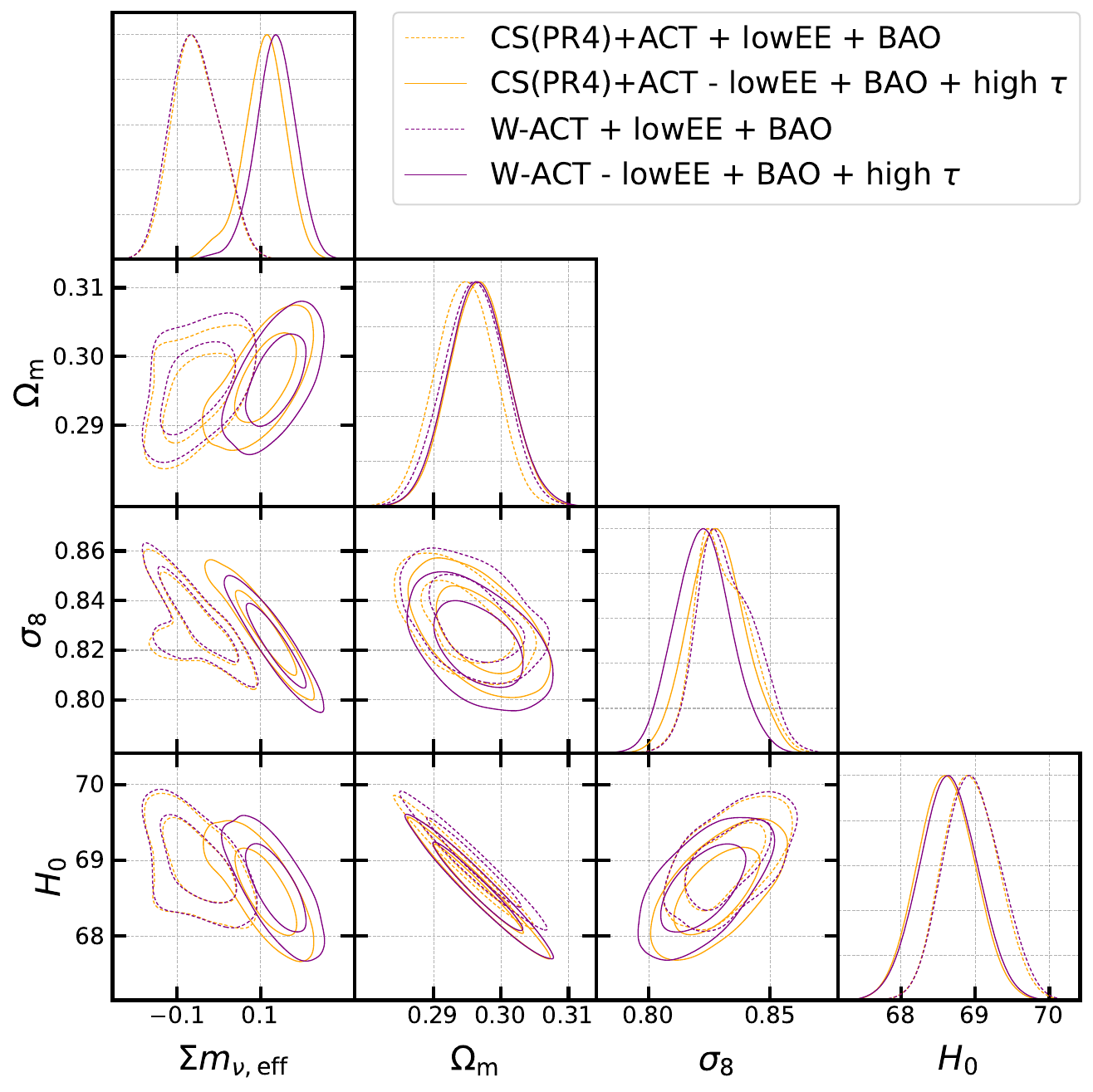}
    \caption{\textbf{Full pairwise posterior contours for lensing anomaly and ``negative'' neutrino masses in 
    alterative CMB likelihoods:} Contours have the same meaning as in Fig.~\ref{fig:AL_mnu_Omm}.
    \textit{Left}: 
    The free-$A_\mathrm{L}$ posteriors illustrate that the $\Lambda$CDM parameters shown are all correlated with the free lensing amplitude $A_\mathrm{L}$ and that the high-$\tau$ prior permits the points consistent with no lensing anomaly to shift to lower $\Omega_m$, higher $H_0$, and higher $\sigma_8$.
    \textit{Right}: 
    The ``negative'' effective neutrino mass model results.
    As in Fig.~\ref{fig:negnumass_1d}, the $\sum m_{\nu,\mathrm{eff}}$ contours include DESI DR2 data.
    The unphysical preference for ``negative'' effective neutrino mass persists regardless of the alternative likelihood used, though this preference.
    Notably, the allowed range of the degenerate combination of $\sum m_{\nu,\mathrm{eff}}$ and $\Omega_m$ shifts to a range where physical neutrino masses coincide with the $\Omega_m$ preferred by DESI DR2 BAO.
    }
    \label{fig:AL_mnu_triangle}
\end{figure*}

Figure~\ref{fig:AL_mnu_triangle} shows the full set of 1D and 2D marginal posterior constraints discussed in Section~\ref{subsec:cmb}.
We also provide tabulated 1D marginal posterior constraints for the results displayed in Fig.~\ref{fig:alens_1d} and Fig.~\ref{fig:negnumass_1d} in Table~\ref{tab:mnueff_alens}. 

\begin{table*}[t]
\centering
\begin{tabular}{|c|cc|} \hline
    \textbf{Dataset}  
    & $A_{\rm L}$
    & $\sum m_{\nu,{\rm eff}}$
    \\
    \hline
    CS(PR4)+ACT + low-$\ell$ EE
    & $1.0708 \pm 0.0330$
    & --
    \\
    CS(PR4)+ACT - low-$\ell$ EE + h$\tau$
    & $0.9693_{-0.0303}^{+0.0299}$
    & --
    \\
    W+ACT + low-$\ell$ EE
    & $1.0161_{-0.0494}^{+0.0487}$
    & --
    \\
    W+ACT - low-$\ell$ EE + h$\tau$
    & $0.9280_{-0.0490}^{+0.0440}$
    & --
    \\
    CS(PR4)+ACT + low-$\ell$ EE
    & --
    & $-0.170_{-0.135}^{+0.137}$ \\
    CS(PR4)+ACT - low-$\ell$ EE + h$\tau$
    & --
    & $0.227_{-0.114}^{+0.140}$ \\
    W+ACT + low-$\ell$ EE
    & --
    & $0.138_{-0.223}^{+0.214}$ \\
    W+ACT - low-$\ell$ EE + h$\tau$ 
    & --
    & $0.434_{-0.162}^{+0.163}$ \\
    CS(PR4)+ACT + low-$\ell$ EE + BAO 
    & --
    & $-0.05_{-0.06}^{+0.05}$
    \\
    CS(PR4)+ACT - low-$\ell$ EE + h$\tau$ + BAO
    & --
    & $0.11_{-0.04}^{+0.05}$
    \\
    W+ACT + low-$\ell$ EE + BAO 
    & --
    & $-0.05_{-0.07}^{+0.05}$
    \\
    W+ACT - low-$\ell$ EE + h$\tau$ + BAO 
    & --
    & $0.14 \pm 0.05$
    \\
    \hline
\end{tabular}
\caption{\textbf{$A_\mathrm{L}$ and $\sum m_{\nu,\mathrm{eff}}$ constraints:}
1-$\sigma$ constraints on the lensing anomaly parameter $A_\mathrm{L}$ as well as the effective neutrino mass parameter $\sum m_{\nu,\mathrm{eff}}$ (in eV).
In the first column,
CS(PR4)+ACT shows significant preference for a $A_\mathrm{L}>1$ lensing anomaly, while W+ACT shows no such preference.
The addition of the high-$\tau$ prior leads to a ``reversed'' lensing anomaly, which admits physcal neutrino mass.
In the second column, we see that the unphysical preference  for ``negative'' neutrino mass is present for both likelihoods \textit{despite} their differing preferences for $A_\mathrm{L}$ and that this unphysical preference is removed by the high-$\tau$ prior.
}
\label{tab:mnueff_alens}
\end{table*}

\bibliographystyle{aux_files/JHEP}
\bibliography{references}

\end{document}